\documentstyle[amstex,preprint,amssymb,amsbsy,eqsecnum,aps,epsfig,graphicx,multicol,prb]{revtex}

\begin{document}
\draft
\title{Electrodynamics of quasi-periodic media}
\author{V. V. Skadorov$^{a,b}$, S. I. Tiutiunnikov$^{a}$}
\address{$^{a}$Joint Institute for Nuclear Research (JINR), Particle Physics
\\
Laboratory,\\
Dubna, 141980, Russia. \\
$^{b}$Institute of Nuclear Problems, Belarus State University,\\
Bobruiskaya str. 11, Minsk, 220050, Belarus .}
\date{\today}
\maketitle

\begin{abstract}
In this work the interaction of electromagnetic field with quasi-periodic
media has been scrutinized. We have obtained the formula for a distorted
medium polarizability tenzor in the X-ray frequency band. Also there have
been obtained the X-rays dynamic diffraction equations for the mediums with
arbitrary smooth distortion. For these equations the Takagi-Tuipin equations
are obtained as a particular case. We have got a very simple formula for the
coefficient of X-rays reflection on the bent Bragg mirror. We have
scrutinized the example of calculating the X-rays reflection and focusing by
the bent Bragg mirror. Also in this work one could find the rezults of
calculating the reflected X-rays caustics form. It is shown that
there exists a possibility of linear dimensions of the X-rays focusing area
having typical values of $10^{-7}cm$.
\end{abstract}

\everymath{\displaystyle}

\section{ INTRODUCTION}

\bigskip

\bigskip

In this article we consider interaction of electromagnetic field with
quasi-periodic media which we define as media obtained by some smooth
deformation of perfect periodic structures. Mathematically, the term
''smooth deformation'' means the biunivocal smooth mapping of the medium
under consideration on the medium with perfect periodic structure. The main
goal of this article is to show that the electrodynamics of quasi-periodic
media can be reduced to the electrodynamics of the perfect periodic media,
and, thus, to make possible application to quasi-periodical structures of
well-developed classical approaches.The correspondence between perfect \ and
quasi-periodical media is possible when the later one is considered as the
Riemannian manifold with the fundamental metric tensor $g_{\alpha \beta }=$ $
\delta _{\alpha \beta }+2\varepsilon _{\alpha \beta }$, where $\delta
_{\alpha \beta }$ is the Kronecker symbol and $\varepsilon _{\alpha \beta }$
is the deformation tensor. To clarify this statement, let us remind, that in
the mechanics of continuous medium there are two methods describing the
deformation processes: the Euler method and the Lagrange method. The
co-ordinates coincide with the observer's frame of reference in the Euler
method. The method of Lagrange stands that the co-ordinates are mounted into
a medium and are deformed together with a medium. It should be emphasized
that the co-ordinates of some point of a medium in the Lagrange co-ordinate
system remain unchanged during the medium deformation, whereas the
co-ordinates of some point of the medium in the Euler co-ordinate system are
being changed. In particular, the co-ordinates of Bravais lattice points of
a medium with perfect periodic structure, defined before the deformation by
a set of three integers ${\bf n}=\{n_{1},n_{2},n_{3}\}$ will remain constant
both during the deformation and after the deformation. Hence, the Bravais
lattice of a deformed (quasi-periodic) medium will look alike in the
Lagrange and the Euler co-ordinate systems. However, it is necessary to
remember, that the area engaged in a quasi-periodic medium in the Lagrange
co-ordinate system is a Riemannian manifold with the fundamental metric
tensor varying from point to point.

Let us give some examples how the formalism developed in this article can be
used while describing the interaction of electromagnetic field with media.
Generally even perfect crystals can be referred rather to the quasi-periodic
media, than to the media with perfect periodic structure because there are
always temperature fields of deformation at least. The classic and quantum
superlattices electrodynamics of semiconductors also can be referred to the
electrodynamics of the quasi-periodic media when it is necessary to take
into account the fields of the deformation, permanently existing in such
structures. A big amount of articles devoted to the electrodynamics of the
carbonic nanotubes is now published. In most cases they take into the
consideration the perfectly straight nanotubes with the perfect periodic
structure. Actually there are always some kinds of the nanotube deformation.
It is possible to reduce the electrodynamics of such nanotubes, and
furthermore the electrodynamics of ensemble of nanotubes in some matrix
completely to the electrodynamics of the quasi-periodic media. In real
crystals the ensemble of the flaws, dislocations and impurities creates a
field of deformations, which is usually divided on two parts. The first one
is the averaging over the ensemble and is the smooth field of deformations.
The second one describes the fluctuations of the deformations field in
comparison with this smooth averaged field. The interaction of an
electromagnetic field with the crystals, that have the deformation field
being the averaged smooth field of deformations, can be described by methods
suggested in this article.

The last example to be considered is the X-ray diffraction optics of the
elastically deformed perfect crystals. Such crystals completely correspond
to the definition of quasi-periodic media given above. From the point of
view of the classical electrodynamics, the X-ray diffraction optics is a
simple version of the geometrical optics. The only complication are coupled
waves inside a crystal which appear because of the dynamic diffraction.
Despite the distorted Bragg mirror is one of the most important parts of
almost all X-ray optics system, we have not seen papers correctly describing
a dynamic X-rays diffraction on elastically deformed crystals with mean
radius of curvature less or about $10m$ whereas elastically deformed
crystals with such media radiuses of curvature are of practical interest.
Takagi-Taupin equations \cite{[1]}-\cite{[2]} and the series of articles
(see bibliography), using this equations as the basic ones are valid for
elastically deformed crystals with mean radius of curvature more than $40m$,
as will be shown in part 4. Furthermore, these papers and also paper \cite%
{[5]}, do not show the difference between two methods describing the crystal
deformation mentioned above (Euler and Lagrange methods). The fact is also
ignored that the deformed crystal in the Lagrange co-ordinate system is the
Riemannian manifold with a fundamental metric tensor varying from point to
point in the crystal. Such disregardings lead to a paradoxical statement,
that it is impossible to destinguish a perfect crystal and the deformed
crystal in the experiments on X-ray beam passing through a crystal far away
from the Bragg condition. That contradicts both daily practice of X-ray
beams application and classical electrodynamics because the deformed crystal
is a nonuniform medium with refraction index and absorption coefficient
depending on co-ordinates.

The first section of this article is dedicated to the electrodynamics of
media with perfect periodic structure and is being auxiliary. As the medium
with periodic structure is a medium with a spatial dispersion and the
relation between the vector of polarization and the electric vector is not
local, the wave equation is the integro-differential one and is rather
difficult to work with. In case of media with perfect periodic structure one
usually proceeds to the ($k,\omega $)-representation, where the
integro-differential wave equation is replaced by the system of the
algebraic equations. Then one derives the medium eigenmodes from the
dispersion equation and gains all key results of the electrodynamics of
media with perfect periodic structure. Such an algorithm defies the
quasi-periodic media generalization, therefore in the first section we
expound an unknown method, unwieldy and inconvenient on the first sight,
that allows to substitute the integro-differential wave equation for the
equivalent system of the differential equations, that describes the
electromagnetic field dynamic diffraction by the media with perfect periodic
structure. This method can be naturally generalized for the case of the
quasi-periodic media.

The second section is devoted to the electrodynamics of quasi-periodic
media. Proceeding to the Lagrange co-ordinate system allows us to apply the
formalism of media with perfect periodic structure developed in the first
section to the quasi-periodic media almost without changes in case we
consider a quasi-periodic medium in a Lagrange co-ordinate system being a
Riemannian manifold.

In the third section we obtain the polarization tensor of a deformed crystal
in the X-ray frequency band from the first principles by the joint solution
of Maxwell equations and von Neumann's equation for a density matrix of a
crystal. It is necessary to underline here, that the dependence of the
Debye-Waller factor on the co-ordinates is ignored, the Fourier-components
of the polarization tensor of a deformed crystal coincide with
Fourier-components of the polarization tensor of a perfect crystal by the
shape only in a Lagrange co-ordinate system. These components appear to be
very complicated co-ordinate functions in the observer system (Euler
co-ordinate system).

In the fourth section we obtain the equations system of the two-wave X-rays
diffraction on the deformed crystals and after a series of simplifying
approximations we show that it is possible to receive from these equations a
combined Takagi -Taupin \cite{[1]}-\cite{[2]} equations system which appears
applicable only for the deformations of the crystals with medial radius of
curvature more than $40m$ .

Finally in the fifth section we use a principle of the geometrical optics
locality to offer a simple formula for the Bragg reflection coefficient from
distorted Bragg mirror and describe the X-ray quanta mirror focusing.

\section{ Perfect periodic structures}

First we consider interaction of the electromagnetic field with a perfect
periodic medium. For media with spatial dispersion, the wave equation for
electric field vector in $\omega $-representation has the form as follows:
\begin{equation}
rotrot{\bf E}({\bf r},\omega )=\frac{\omega ^{2}}{c^{2}}{\bf E}({\bf r}
,\omega )+\frac{\omega ^{2}}{c^{2}}\int d^{3}{\bf r}^{\prime }\hat{\chi}(
{\bf r},{\bf r}^{\prime };\omega ){\bf E}({\bf r}^{\prime },\omega )+\frac{
4\pi \omega i}{c^{2}}{\bf j}^{(e)}({\bf r},\omega )  \label{1}
\end{equation}
where $\hat{\chi}({\bf r},{\bf r}^{\prime };\omega )$ is the medium
polarization tensor, and ${\bf j}^{(e)}({\bf r},\omega )$ is the current
density of external sources. The intrinsic property of perfect periodic
media to be translationally invariant, makes it possible to expand the
polarizability tensor into the Fourier series:

\begin{equation}
\hat{\chi}({\bf r},{\bf r}^{\prime };\omega )=\hat{\chi}({\bf r}+{\bf R}_{%
{\bf n}},{\bf r}^{\prime }+{\bf R}_{{\bf n}};\omega )=\hat{\chi}({\bf r}-
{\bf r}^{\prime },{\bf r}^{\prime };\omega )=\sum\limits_{{\bf l}}\hat{\chi}
^{{\bf l}}({\bf r}-{\bf r}^{\prime };\omega )\exp \left\{ -i\left\langle
{\bf \tau }^{({\bf l})},{\bf r}^{\prime }\right\rangle \right\}  \label{2}
\end{equation}
In the above equations ${\bf R}_{{\bf n}}=n_{1}{\bf a}_{1}+n_{2}{\bf a}
_{2}+n_{3}{\bf a}_{3}$ is the Bravais lattice vector, ${\bf n}
=\{n_{1},n_{2},n_{3}\}$ and ${\bf l}=\{l_{1},l_{2},l_{3}\}$ are integer
quantities, ${\bf \tau }^{({\bf l})}=l_{1}{\bf b}_{1}+l_{2}{\bf b}_{2}+l_{3}
{\bf b}_{3}$ are the reciprocal lattice vectors, ${\bf a}_{1}$, ${\bf a}_{2}$
, ${\bf a}_{3}$ and ${\bf b}_{1},{\bf b}_{2},{\bf b}_{3}$ are the basic
vectors of the Bravais and reciprocal lattices, correspondingly.

Let an electromagnetic wave be incident on a periodic medium with a given
Bravais lattice. If the wavelength of the incident electromagnetic field is
of the order of the elementary sell linear extension, solution of Eq. (\ref%
{1}) is reasonable to seek in the form of the expansion

\begin{equation}
{\bf E}({\bf r},\omega )=\sum\limits_{{\bf l}}{\bf E}_{{\bf l}}({\bf r}
,\omega )e^{i({\bf k}+{\bf \tau }^{({\bf l})}){\bf r}}  \label{3}
\end{equation}
where the amplitudes ${\bf E}_{{\bf l}}({\bf r},\omega )$ vary slowly in the
area with the linear dimensions less then the wavelength and $({\bf k})^{2}=
\frac{\omega ^{2}}{c^{2}}$. By substitution Eqs. (\ref{2})-(\ref{3}) into
Eq. (\ref{1}) we obtain

\begin{equation}
\sum\limits_{{\bf l}}e^{i{\bf \tau }^{({\bf l})}{\bf r}}\{\hat{L}[({\bf k}+
{\bf \tau }^{({\bf l})}),\nabla ]{\bf E}_{{\bf l}}({\bf r},\omega
)+k^{2}\sum\limits_{{\bf l}^{\prime }}\int d^{3}{\bf r}^{\prime }\hat{\chi}^{%
{\bf l}^{\prime }}({\bf r}^{\prime },\omega )e^{-i({\bf k}+{\bf \tau }^{(
{\bf l})}){\bf r}^{\prime }}{\bf E}_{{\bf l+l}^{\prime }}({\bf r}-{\bf r}
^{\prime },\omega )=0  \label{4}
\end{equation}
where ${\bf j}^{(e)}({\bf r},\omega )=0$ and the differentiation operator $
\hat{L}[({\bf k}+{\bf \tau }^{({\bf l})}),\nabla ]$ is determined by
\[
\hat{L}[({\bf k}+{\bf \tau }^{({\bf l})}),\nabla ]=\left[ \nabla +i\left(
{\bf k}+{\bf \tau }^{({\bf l})}\right) \right] \times \left[ \nabla +i\left(
{\bf k}+{\bf \tau }^{({\bf l})}\right) \right] .
\]
The symbol $\times $ stands for the vector product. Expanding the amplitude $
{\bf E}_{{\bf l+l}^{\prime }}({\bf r}-{\bf r}^{\prime },\omega )$ into the
Taylor series and substituting this series into the integral in expression ( %
\ref{4}) we obtain

$\int d^{3}{\bf r}^{\prime }\hat{\chi}^{{\bf l}^{\prime }}({\bf r}^{\prime
},\omega )e^{-i({\bf k}+{\bf \tau}^{({\bf l})}){\bf r}^{^{\prime }}}{\bf E}_{%
{\bf l+l}^{\prime }}({\bf r}-{\bf r}^{\prime },\omega )=\hat{\chi}^{{\bf l}
^{\prime }}({\bf k}+{\bf \tau}^{({\bf l})},\omega ){\bf E}_{{\bf l+l}
^{\prime }}({\bf r},\omega )+$

$i {\displaystyle{\frac{\partial }{\partial k_{i}}}} \hat{\chi}^{{\bf l}%
^{\prime }}({\bf k}+{\bf \tau}^{({\bf l})},\omega ) {\displaystyle{\frac{%
\partial {\bf E}_{{\bf l+l}^{\prime }}({\bf r},\omega ) }{\partial x^{i}}}} -%
\frac{1}{2} {\displaystyle{\frac{\partial }{\partial k_{i}\partial k_{j}}}}
\hat{\chi}^{{\bf l}^{\prime }}({\bf k}+{\bf \tau}^{({\bf l})},\omega ) {%
\displaystyle{\frac{\partial ^{2}{\bf E}_{{\bf l+l}^{\prime }}({\bf r}%
,\omega ) }{\partial x^{i}\partial x^{j}}}} +\cdots $.

Here and below, the summation over repetitive indices is implicit. Since $
\chi _{\alpha ,\beta }^{{\bf ll^{\prime }}}=O(1/k^{2})$, the estimate $[\chi
_{\alpha ,\beta }^{{\bf ll^{\prime }}}]^{-1}\partial \chi _{\alpha ,\beta }^{%
{\bf ll^{\prime }}}/\partial k_{i}=O(1/k)$ holds true and, as a result, $
\partial \chi _{\alpha ,\beta }^{{\bf ll^{\prime }}}/\partial k_{i}=o(\chi
_{\alpha ,\beta }^{{\bf ll^{\prime }}})$. Then, in view of that the
amplitudes ${\bf E}_{{\bf l+l}^{\prime }}({\bf r},\omega )$ are slowly
varying functions, all terms in the last expression except the first one can
be omitted and, consequently, integro-differential equation (\ref{4})
reduces to

\begin{equation}
\sum\limits_{{\bf l}}e^{i{\bf \tau }^{({\bf l})}{\bf r}}\{\hat{L}[({\bf k}+
{\bf \tau }^{({\bf l})}),\nabla ]{\bf E}_{{\bf l}}({\bf r},\omega
)+k^{2}\sum\limits_{{\bf l}^{\prime }}\hat{\chi}^{{\bf l}^{\prime }}({\bf k}%
+ {\bf \tau }^{({\bf l})},\omega ){\bf E}_{{\bf l+l}^{\prime }}({\bf r}%
,\omega )\}=0.  \label{5}
\end{equation}
Let ${\bf r}_{0}$ be the radius-vector of the elementary sell center. In
that case, any point ${\bf r}$ inside the sell is given by ${\bf r}={\bf r}
_{0}+{\bf \eta }$. Taking this representation into account, let us multiply
the last equation by $\exp \{-i{\bf \tau }^{({\bf l})}{\bf r}\}$ and
integrate the product over the elementary sell volume $\Omega $:

\begin{eqnarray}
&&\sum\limits_{{\bf l}^{\prime }}e^{i({\bf \tau }({\bf l}^{\prime })-{\bf %
\tau }^{({\bf l})}){\bf r}_{0}}\{\int\limits_{\Omega }e^{i({\bf \tau }({\bf %
l }^{\prime })-{\bf \tau }^{({\bf l})}){\bf \eta }}\hat{L}[({\bf k}+{\bf %
\tau } ({\bf l}^{\prime })),\nabla ]{\bf E}_{{\bf l}^{\prime }}({\bf r}_{0}+%
{\bf \eta },\omega )d^{3}{\bf \eta } \\
&&+k^{2}\sum\limits_{{\bf l}^{^{\prime \prime }}}\hat{\chi}^{{\bf l}
^{^{\prime \prime }}}({\bf k}+{\bf \tau }({\bf l}^{\prime }),\omega
)\int\limits_{\Omega }e^{i({\bf \tau }({\bf l}^{\prime })-{\bf \tau }^{({\bf %
l})}){\bf \eta }}{\bf E}_{{\bf l}^{\prime }{\bf +l}^{^{\prime \prime }}}(
{\bf r}_{0}+{\bf \eta },\omega )d^{3}{\bf \eta }\}=0
\end{eqnarray}
In the framework of the slowly varying amplitude approximation taken above,
the quantities $\hat{L}[({\bf k}+{\bf \tau }(l^{^{\prime }})),\nabla ]{\bf E}
_{{\bf l}^{\prime }}({\bf r}_{0}+{\bf \eta },\omega )$ and${\bf E}_{{\bf l}
^{\prime }{\bf +l}^{^{\prime \prime }}}({\bf r}_{0}+{\bf \eta },\omega )$
can be treated as constants within the elementary sell and, thus, can be
factored out from the integrals. In that case, since

\[
\ \int\limits_{\Omega }e^{i({\bf \tau }({\bf l}^{\prime })-{\bf \tau }^{(
{\bf l})}){\bf \eta }}d^{3}{\bf \eta }=\Omega \delta _{{\bf l,l}^{\prime }}
\]
the equation (\ref{5}) reduces to the system as follows:

\begin{equation}
\hat{L}[({\bf k}+{\bf \tau }^{({\bf l})}),\nabla ]{\bf E}_{{\bf l}}({\bf r}
,\omega )+k^{2}\sum\limits_{{\bf l}^{\prime }}\hat{\chi}^{{\bf l}^{\prime
}}( {\bf k}+{\bf \tau }^{({\bf l})},\omega ){\bf E}_{{\bf l+l}^{\prime }}(%
{\bf r} ,\omega )=0  \label{6}
\end{equation}
Together with Eq. (\ref{3}), this equation system describes the
electromagnetic field dynamic diffraction by a perfect periodic medium.

When the wavelength is much larger then the medium unit cell linear size,
electromagnetic fields propagating in the medium can be averaged over the
elementary sell and, then, a hypotatic homogeneous medium with effective
constitutive parameters can be introduced into consideration instead of the
real inhomogeneous one. Polarization of this homogeneous medium is related
to averaged electromagnetic field by ${\bf P}_{avg}({\bf r},\omega )=\hat{
\chi}_{avg}{\bf E}_{avg}({\bf r},\omega )$, where $\hat{\chi}_{avg}$ is the
polarization tensor which determines the linear electromagnetic response of
the effective homogeneous medium. \ In accordance with \cite{[7]}, the
effective polarizability tensor $\hat{\chi}^{avg}({\bf k};\omega )$ is
expressed in terms of the Fuorier transform of the microscopic
polarizability tensor (\ref{2}) by

\bigskip
\begin{equation}
\hat{\chi}^{avg}({\bf k};\omega )=\hat{\chi}^{{\bf 0}}({\bf k};\omega )\frac{
1}{1+\hat{N}\hat{\chi}^{{\bf 0}}({\bf k};\omega )}  \label{7}
\end{equation}

The tensor $\hat{N}$ \ takes into account interaction between scattering
centers and is determined by the lattice geometry:

\begin{eqnarray}
\hat{N} &=&\lim_{{\bf r}\rightarrow {\bf R}_{{\bf m}}}({\rm graddiv}
+\varkappa ^{2})\int\limits_{\Omega }d^{3}{\bf \eta }^{\prime }\frac{
e^{-i\varkappa \mid {\bf r}-{\bf R}_{{\bf m}}-{\bf \eta }^{\prime }\mid }}{
\mid {\bf r}-{\bf R}_{{\bf m}}-{\bf \eta }^{\prime }\mid }\varphi ({\bf \eta
}^{^{\prime }})\thickapprox  \nonumber \\
&&\lim\limits_{{\bf r}\rightarrow {\bf R}_{{\bf m}}}{\rm graddiv}
\int\limits_{\Omega }d^{3}{\bf \eta }^{\prime }\frac{\varphi ({\bf \eta }
^{\prime })}{\mid {\bf r}-{\bf R}_{{\bf m}}-{\bf \eta }^{\prime }\mid }
\label{8}
\end{eqnarray}
The function $\varphi ({\bf \eta })$ \ is an even function of co-ordinates.
In centro-symmetrical lattices it is defined by its momenta:

\begin{equation}
\frac{1}{\Omega }\int\limits_{\Omega }d^{3}{\bf \eta }\varphi ({\bf \eta }
)=1,\frac{1}{\Omega }\int\limits_{\Omega }d^{3}{\bf \eta }\eta ^{i}\varphi (
{\bf \eta })=0,\frac{1}{\Omega }\int\limits_{\Omega }d^{3}{\bf \eta }\eta
^{i}\eta ^{j}\varphi ({\bf \eta })=\delta ^{ij},\cdots .  \label{8a}
\end{equation}
The tensor $\hat{N}$ has been evaluated analytically for some types of
elementary cells \cite{[7]}. In other cases it can be done numerically.

\section{ Quasi-periodic structures}

The basic assumption which the procedure of derivation of Eqs. (\ref{6})-( %
\ref{8}) leaned upon, is the relation (\ref{3}) accounting for the property
of the polarizability tensor of perfect periodic structures to be
translationally invariant. Under effect of deformation, this invariance,
e.g. condition (\ref{3}), is disrupted and thus the algorithm of Eqs. (\ref%
{6})-(\ref{7}) derivation outlined above becomes invalid. However, an
appropriate modification based on the approach originated from the
continuous medium mechanics can be applied to involve deformed periodic
media into consideration. In the mechanics of continuous media there are two
alternative methods of the deformation process description, the Euler method
and the Lagrange method. In the framework of the first one the co-ordinate
system is coupled with the viewpoint whereas the second method treats the
co-ordinate system frozen into the medium and, thus, in the process of
deformation the co-ordinate lines are transformed along with the medium. As
a result, the Euler co-ordinates of a given point $x^{i}$ are changed during
the deformation whereas the Lagrange co-ordinates of it, $\xi ^{i}$, remain
fixed. Lagrange and Euler co-ordinates are related to one another by
\begin{equation}
x^{i}=\xi ^{i}+u^{i}(\xi ^{1},\xi ^{2},\xi ^{3});\quad i=1,2,3\,,  \label{9}
\end{equation}
where $u^{i}(\xi ^{1},\xi ^{2},\xi ^{3})$ stand for the deformation field
components. In particular, choosing the co-ordinate system ${\bf e}_{i}$ to
be related to the elementary translation vectors of the Bravais lattice $
{\bf a}_{i}$, ${\bf e}_{i}={\bf a}_{i}/|{\bf a}_{i}|$, we find out the
Lagrange co-ordinates of the Bravais lattice sites to be given by quantities
$\xi _{{\bf n}}=\{n_{1}a_{1},n_{2}a_{2},n_{3}a_{3}\}$ both before and after
the deformation. Thus, one can conclude that in the Lagrange co-ordinates,
the Bravais lattice of a deformed periodic structure looks like to the
Bravais lattice of a perfect periodic structure and consequently, the
polarizability tensor in the Lagrange co-ordinates proves to be
translationally invariant. This allows us to state the relations as follows
\begin{equation}
\hat{\chi}({\bf \xi },{\bf \xi }^{\prime };\omega )=\hat{\chi}({\bf \xi }+
{\bf \xi }_{{\bf n}},{\bf \xi }^{\prime }+{\bf \xi }_{{\bf n}};\omega )=\hat{
\chi}({\bf \xi }-{\bf \xi }^{\prime },{\bf \xi }^{\prime };\omega
)=\sum\limits_{{\bf l}}\hat{\chi}^{{\bf l}}({\bf \xi }-{\bf \xi }^{^{\prime
}};\omega )\exp \left\{ -i\left\langle {\bf \tau }^{({\bf l})},{\bf \xi }
^{\prime }\right\rangle \right\}  \label{10}
\end{equation}
where ${\bf \tau }^{({\bf l})}=\{{2\pi l_{1}/a_{1}},{2\pi l_{2}/a_{2}},{2\pi
l_{3}/a_{3}}\}$ is the reciprocal lattice covector in the Lagrange
co-ordinate system; its components in the Euler co-ordinates are defined by $%
\tau _{{\bf l}i}({\bf r})=(2\pi l_{j}/a_{j}){\partial \xi ^{j}}/{\partial
x^{i}},~i=1,2,3\,.$. It should be emphasized that, mathematically, a
deformed medium is the Riemann manifold with the fundamental metric tensor
as follows
\[
g_{ij}(\xi ^{1},\xi ^{2},\xi ^{3})=g_{ij}^{0}+2\varepsilon _{ij}(\xi
^{1},\xi ^{2},\xi ^{3})
\]
where $\varepsilon _{ij}(\xi ^{1},\xi ^{2},\xi ^{3})$ is the deformation
tensor of the medium:

\begin{equation}
\varepsilon _{ij}(\xi ^{1},\xi ^{2},\xi ^{3})=\frac{1}{2}(g_{ik}^{0}\frac{
\partial u^{k}}{\partial \xi ^{j}}+g_{jk}^{0}\frac{\partial u^{k}}{\partial
\xi ^{i}}+g_{lk}^{0}\frac{\partial u^{l}}{\partial \xi ^{j}}\frac{\partial
u^{k}}{\partial \xi ^{i}})  \label{11}
\end{equation}
Here $g_{ij}^{0}=({\bf e}_{i}\cdot {\bf e}_{j})$ is the metric tensor of
elementary cell of the perfect periodic medium. Further mathematical
analysis accounted for the properties of the Riemann manifolds allows one to
extend the approach being developed to media which need not be obtained by
deformation of the perfect periodic medium. In order for relation (\ref{9})
to hold true, a one-to-one smooth mapping (\ref{10}) of the medium being
considered into a perfect periodic structure is necessary and sufficient to
exist. The media for which such a mapping exists we shall refer to as
quasi-periodic media. Wave equation for quasi-periodic media in the Lagrange
co-ordinates takes the form as follows
\begin{eqnarray}
&&\lbrack e_{h}({\bf \xi })\otimes g^{kh} {\displaystyle{\frac{\partial }{%
\partial \xi ^{k}}}} \frac{1}{\sqrt{g}} {\displaystyle{\frac{\partial }{%
\partial \xi ^{l}}}} \sqrt{g}e^{l}({\bf \xi })-\frac{1}{\sqrt{g}} {%
\displaystyle{\frac{\partial }{\partial \xi ^{l}}}} \sqrt{g}g^{kl} {%
\displaystyle{\frac{\partial }{\partial \xi ^{k}}}} ]{\bf E}({\bf \xi }%
,\omega )=\frac{\omega ^{2}}{c^{2}}{\bf E}({\bf \xi } ,\omega )  \nonumber \\
&&+\frac{\omega ^{2}}{c^{2}}\sum\limits_{{\bf l}}\int d^{3}{\bf \xi }%
^{\prime } \sqrt{\left| g({\bf \xi }^{\prime })\right| }\hat{\chi}^{{\bf l}}(%
{\bf \xi }- {\bf \xi }^{\prime };\omega )e^{-i\tau _{i}^{({\bf l})}\xi ^{i}}%
{\bf E}({\bf \xi }^{\prime },\omega )+\frac{4\pi \omega i}{c^{2}}{\bf j}%
^{(e)}({\bf \xi } ,\omega )  \label{12}
\end{eqnarray}
where $g^{ij}(\xi )$ is the tensor inverse to the metric tensor $g_{ij}(\xi
) $ determined by Eq. (\ref{11})\ \ ($g^{ij}(\xi )g_{jk}(\xi )=\delta
_{k}^{i} $), \ $g(\xi )=\det \left\| g_{ij}({\bf \xi })\right\| .$

The procedure of derivation of basic equation describing wave diffraction by
quasi-periodic media is analogous to that applied above under derivation of
Eqs. (\ref{6})-(\ref{7}). As in the case of perfect periodic media, solution
of wave equation (\ref{12}) we shell seek in the form of expansion analogous
to (\ref{3}):

\begin{equation}
{\bf E}({\bf \xi },\omega )=\sum\limits_{{\bf l}}{\bf E}_{{\bf l}}({\bf \xi }
,\omega )e^{i[k_{i}+\tau _{i}^{({\bf l})}]\xi ^{i}+ik_{i}u^{i}({\bf \xi })}
\label{13}
\end{equation}
\ \ \ where $k_{i}$are the wave vector ${\bf k}$ components in the Euler
co-ordinate system. As before, the wavelength $\lambda =2\pi c/\omega $ is
assumed to be of the order of the elementary cell linear size. It should be
emphasized that expression (\ref{13}), as different from Eq. (\ref{3}), is
not the expansion in terms of plane waves because the phases $\Phi _{{\bf l}
}({\bf k},\xi )=\Phi _{{\bf l}}({\bf k},{\bf r})=(k_{i}+\tau _{i}^{({\bf l}
)})\xi ^{i}+k_{i}u^{i}(\xi )=k_{i}x^{i}+ {\displaystyle{\frac{\partial \xi
^{j}({\bf r}) }{\partial x^{i}}}} \tau _{j}^{({\bf l})}\xi ^{i}({\bf r})$
are intricate functions of co-ordinates in both Euler and Lagrange systems.
Letting in Eq. (\ref{12}) ${\bf j}^{(e)}(\xi ,\omega )=0$ and substituting
Eq.(\ref{13}) into (\ref{12} ), one can obtain
\begin{eqnarray}
&&\sum\limits_{{\bf l}}\{e^{i[k_{i}({\bf \xi })+\tau _{i}^{({\bf l})}]\xi
^{i}}[-{\bf e}_{h}({\bf \xi })\otimes \hat{L}_{l}^{h}({\bf k}({\bf \xi })+
{\bf \tau }^{({\bf l})}, {\displaystyle{\frac{\partial }{\partial {\bf \xi }}%
}} ){\bf e}^{l}({\bf \xi })+\hat{L}({\bf k}({\bf \xi })+{\bf \tau }^{({\bf l}%
)}, {\displaystyle{\frac{\partial }{\partial {\bf \xi }}}} )+k^{2}]{\bf E}_{%
{\bf l}}({\bf \xi },\omega )  \nonumber \\
&&+k^{2}\sum\limits_{{\bf l}^{\prime }}\int d^{3}{\bf \xi }^{\prime }\sqrt{%
\left| g({\bf \xi -\xi }^{\prime })\right| }\hat{\chi}^{{\bf l}^{\prime }}(
{\bf \xi }^{\prime };\omega )e^{i[k_{i}+\tau _{i}^{({\bf l})}](\xi ^{i}-\xi
^{^{\prime }i})+ik_{i}u^{i}({\bf \xi -\xi }^{\prime })}{\bf E}_{{\bf l+l}
^{\prime }}({\bf \xi -\xi }^{\prime },\omega )\}=0  \label{14}
\end{eqnarray}
where $k_{i}(\xi )=k_{i}+k_{j} {\displaystyle{\frac{\partial u^{j}({\bf \xi )%
} }{\partial \xi ^{i}}}} $ are the wave vector ${\bf k}$ components in the
Lagrange co-ordinate system, $k(\xi )+\tau ^{({\bf l})}=(k_{1}(\xi )+\tau
_{1}^{({\bf l} )},k_{2}(\xi )+\tau _{2}^{({\bf l})},k_{3}(\xi )+\tau _{3}^{(%
{\bf l})})$, ${\displaystyle{\frac{\partial }{\partial {\bf \xi }}}} =( {%
\displaystyle{\frac{\partial }{\partial \xi ^{1}}}} , {\displaystyle{\frac{%
\partial }{\partial \xi ^{2}}}} , {\displaystyle{\frac{\partial }{\partial
\xi ^{3}}}} )$ and differentiation operators $\hat{L}_{l}^{h}(k(\xi )+\tau
^{({\bf l})}, {\displaystyle{\frac{\partial }{\partial {\bf \xi }}}} )$, $%
\hat{L}(k(\xi )+\tau ^{({\bf l})}, {\displaystyle{\frac{\partial }{\partial
{\bf \xi }}}} )$ are determined by the relations as follows:

\begin{eqnarray}
&&\hat{L}_{l}^{h}({\bf k}({\bf \xi })+{\bf \tau }^{({\bf l})}, {\displaystyle%
{\frac{\partial }{\partial {\bf \xi }}}} )=g^{hj}\{ {\displaystyle{\frac{%
\partial }{\partial \xi ^{j}}}} +i[k_{j}({\bf \xi })+\tau _{j}^{({\bf l})}]\}%
\frac{1}{\sqrt{g}}\{ {\displaystyle{\frac{\partial }{\partial \xi ^{l}}}}
+i[k_{l}({\bf \xi })+\tau _{l}^{({\bf l})}]\}\sqrt{g}  \nonumber \\
&&\hat{L}({\bf k}({\bf \xi })+{\bf \tau }^{({\bf l})}, {\displaystyle{\frac{%
\partial }{\partial {\bf \xi }}}} )=\frac{1}{\sqrt{g}}\{ {\displaystyle{%
\frac{\partial }{\partial \xi ^{l}}}} +i[k_{l}({\bf \xi })+\tau _{l}^{({\bf l%
})}]\}\sqrt{g}g^{lj}\{ {\displaystyle{\frac{\partial }{\partial \xi ^{j}}}}
+i[k_{j}({\bf \xi })+\tau _{j}^{({\bf l})}]\}  \label{14a}
\end{eqnarray}
Integral in Eq. (\ref{14}) can be estimated in the following way. As has
been pointed out above, the polarizability tensor $\hat{\chi}^{{\bf l}
^{\prime }}({\bf k};\omega )$ is proportional to the amplitude of scattering
of electromagnetic field by elementary cell. \ By this reason, the quantity $
\hat{\chi}^{{\bf l}^{\prime }}({\bf k};\omega )$ quickly fall down with $\xi
^{\prime }$ and becomes negligible at distances exceeding linear size of the
elementary cell. It means that the Taylor expansion of $u^{i}(\xi -\xi
^{\prime })=u^{i}(\xi )+\xi ^{^{\prime }j} {\displaystyle{\frac{\partial
u^{i} }{\partial \xi ^{j}}}} +\cdots $ in the vicinity of the point ${\xi }$
in the integrand of Eq. (\ref{14}) can be truncated beyond the first term.
Further analysis allows one to reduce Eq. (\ref{14}) to the system as
follows:

\begin{eqnarray}
&&\int d^{3}{\bf \xi }^{\prime }\sqrt{\left| g({\bf \xi -\xi }^{^{\prime
}})\right| }\hat{\chi}^{{\bf l}^{\prime }}({\bf \xi }^{\prime };\omega
)e^{-i[k_{i}({\bf \xi })+\tau _{i}^{({\bf l})}]\xi ^{^{\prime }i}}{\bf E}_{%
{\bf l+l}^{\prime }}({\bf \xi -\xi }^{\prime },\omega )=  \nonumber \\
&&\sqrt{\left| g({\bf \xi })\right| }\hat{\chi}^{{\bf l}^{\prime }}({\bf k}
^{\prime }({\bf \xi })+{\bf \tau }^{({\bf l})};\omega ){\bf E}_{{\bf l+l}
^{\prime }}({\bf \xi },\omega )+i\sqrt{\left| g({\bf \xi })\right| }\frac{
\partial }{\partial k_{i}}\hat{\chi}^{{\bf l}^{\prime }}({\bf k}({\bf \xi }%
)+ {\bf \tau }^{({\bf l})};\omega )\frac{\partial {\bf E}_{{\bf l+l}^{\prime
}}( {\bf \xi },\omega )}{\partial \xi ^{i}}  \nonumber \\
&&+i\frac{\partial \sqrt{\left| g({\bf \xi })\right| }}{\partial \xi ^{i}}
\frac{\partial }{\partial k_{i}}\hat{\chi}^{{\bf l}^{\prime }}({\bf k}({\bf %
\xi })+{\bf \tau }^{({\bf l})};\omega ){\bf E}_{{\bf l+l}^{\prime }}({\bf %
\xi },\omega )+\cdots  \nonumber
\end{eqnarray}
Hence the integro-differential equation (\ref{14}) can be approximated with
the good accurency by the differential equation

\begin{eqnarray}
&&\sum\limits_{{\bf l}}\{e^{i\tau _{i}^{({\bf l})}\xi ^{i}}\{[-e_{h}({\bf %
\xi })\otimes \hat{L}_{l}^{h}({\bf k}({\bf \xi })+{\bf \tau }^{({\bf l})}, {%
\displaystyle{\frac{\partial }{\partial {\bf \xi }}}} )e^{l}({\bf \xi })+%
\hat{L}({\bf k}({\bf \xi })+{\bf \tau }^{({\bf l})}, {\displaystyle{\frac{%
\partial }{\partial {\bf \xi }}}} )+k^{2}]{\bf E}_{{\bf l}}({\bf \xi }%
,\omega )  \nonumber \\
&&+k^{2}\sum\limits_{{\bf l}^{\prime }}\sqrt{\left| g({\bf \xi })\right| }
\hat{\chi}^{{\bf l}^{\prime }}({\bf k}({\bf \xi })+{\bf \tau }^{({\bf l}
)};\omega ){\bf E}_{{\bf l+l}^{\prime }}({\bf \xi },\omega )\}=0  \nonumber
\end{eqnarray}
Just as for the perfect periodic media it can be proved that this equation
is equivalent to the system

\begin{eqnarray}
&&[-e_{h}({\bf \xi })\otimes \hat{L}_{l}^{h}({\bf k}({\bf \xi })+{\bf \tau }
^{({\bf l})}, {\displaystyle{\frac{\partial }{\partial {\bf \xi }}}} )e^{l}(%
{\bf \xi })+\hat{L}({\bf k}({\bf \xi })+{\bf \tau }^{({\bf l})}, {%
\displaystyle{\frac{\partial }{\partial {\bf \xi }}}} )+k^{2}]{\bf E}_{{\bf l%
}}({\bf \xi },\omega )+  \nonumber \\
&&+k^{2}\sum\limits_{{\bf l}^{\prime }}\sqrt{\left| g({\bf \xi })\right| }
\hat{\chi}^{{\bf l}^{\prime }}({\bf k}({\bf \xi })+{\bf \tau }^{({\bf l}
)};\omega ){\bf E}_{{\bf l+l}^{\prime }}({\bf \xi },\omega )\}=0.  \label{15}
\end{eqnarray}
which describes dynamical diffraction of electromagnetic field by
quasi-periodical media. Differentiation operators $\hat{L}_{l}^{h}(k(\xi
)+\tau ^{({\bf l})}, {\displaystyle{\frac{\partial }{\partial {\bf \xi }}}}
) $, $\hat{L}(k(\xi )+\tau ^{({\bf l})}, {\displaystyle{\frac{\partial }{%
\partial {\bf \xi }}}} )$ \ are determined by the formulas (\ref{14a}).

If the wavelength of the incident electromagnetic field is much larger than
the linear size of the lattice unit cell, the averaging procedure in a
quasi-periodic medium in the Lagrange co-ordinate system is identical to
such procedure for perfect periodic media and leads to the equation

\begin{equation}
\hat{\chi}^{avg}({\bf \xi },{\bf k}({\bf \xi });\omega )=\sqrt{\left| g({\bf %
\xi })\right| }\hat{\chi}^{{\bf 0}}({\bf k}({\bf \xi });\omega )\frac{1}{1+
\sqrt{\left| g({\bf \xi })\right| }\hat{N}({\bf \xi })\hat{\chi}^{{\bf 0}}(
{\bf k}^{\prime }({\bf \xi });\omega )}.  \label{16}
\end{equation}
with the depolarization tensor $\hat{N}(\xi )$ determined by

\begin{eqnarray}
&&\hat{N}({\bf \xi })=\lim_{{\bf \xi }\rightarrow {\bf \xi }^{^{\prime
}}}e_{h}({\bf \xi })\otimes \hat{L}_{l}^{h}({\bf k}({\bf \xi })+{\bf \tau }(
{\bf l}), {\displaystyle{\frac{\partial }{\partial {\bf \xi }}}} )  \nonumber
\\
&&\times \int\limits_{\Omega }d^{3}{\bf \eta }\sqrt{\left| g({\bf \xi }
^{\prime }+{\bf \eta })\right| }\frac{e^{-ik\mid {\bf \xi -\xi }^{\prime }+
{\bf u}({\bf \xi )-}{\bf u}({\bf \xi }^{\prime }+{\bf \eta )}\mid }}{\mid
{\bf \xi -\xi }^{\prime }+{\bf u}({\bf \xi )-}{\bf u}({\bf \xi }^{\prime }+
{\bf \eta )-\eta }\mid }\varphi ({\bf \eta })e^{l}({\bf \xi })\thickapprox \\
&&\lim\limits_{{\bf \xi }\rightarrow {\bf \xi }^{\prime }}e_{h}({\bf \xi }
)\otimes \hat{L}_{l}^{h}({\bf k}({\bf \xi })+{\bf \tau }({\bf l}), {%
\displaystyle{\frac{\partial }{\partial {\bf \xi }}}} )\int\limits_{\Omega
}d^{3}{\bf \eta }\sqrt{\left| g({\bf \xi }^{\prime }+ {\bf \eta })\right| }%
\frac{\varphi ({\bf \eta })}{\mid {\bf \xi -\xi } ^{\prime }+{\bf u}({\bf %
\xi )-}{\bf u}({\bf \xi }^{\prime }+{\bf \eta )-\eta }\mid }e^{l}({\bf \xi }%
),  \nonumber
\end{eqnarray}
where the integration is over unit cell of the perfect periodic medium and
the function $\varphi (\eta )$ is determined by the equations (\ref{8a}).
The equations (\ref{16}-\ref{17}) together with the equation

\begin{eqnarray}
&&\lbrack e_{h}({\bf \xi })\otimes g^{kh} {\displaystyle{\frac{\partial }{%
\partial \xi ^{k}}}} \frac{1}{\sqrt{g}} {\displaystyle{\frac{\partial }{%
\partial \xi ^{l}}}} \sqrt{g}e^{l}({\bf \xi })-\frac{1}{\sqrt{g}} {%
\displaystyle{\frac{\partial }{\partial \xi ^{l}}}} \sqrt{g}g^{kl} {%
\displaystyle{\frac{\partial }{\partial \xi ^{k}}}} ]{\bf E}({\bf \xi }%
,\omega )=  \nonumber \\
&&\frac{\omega ^{2}}{c^{2}}{\bf E}({\bf \xi },\omega )+\frac{\omega ^{2}}{
c^{2}}\hat{\chi}^{avg}({\bf \xi },{\bf k}({\bf \xi });\omega ){\bf E}({\bf %
\xi },\omega )+\frac{4\pi \omega i}{c^{2}}{\bf j}^{(e)}({\bf \xi },\omega )
\label{18}
\end{eqnarray}
are determining the electrodynamics of the quasi-periodic medium in this
frequency band. \ \

\ \ \ \ \ \ \ \ \ \ \ \ \ \ \ \

\section{ The polarizability tensor of elastically deformed crystals in the
x-ray range}

In previous section we have obtained the basic equations of electrodynamics
of quasi-periodic media. Now, starting with the Maxwell equations and von
Neumann equation for the density matrix, we derive an explicit expression
for the polarizability tensor in deformed crystals, which constitute a
significant class of quasi-periodic media. Since the crystal unit cell is
about a few Angstroms in size, dynamical diffraction by crystals manifests
itself in the X-ray frequency range. Below we restrict consideration to this
frequency range. As has been shown above, a deformed crystal in the Lagrange
co-ordinate system is similar to a perfect crystal in the Euler system;
consequently, the Lagrange co-ordinate system is preferable for evaluation
of the polarizability tensor of deformed crystals.

The current density induced in the point $\xi =(\xi ^{1},\xi ^{2},\xi ^{3})$
by X-ray quanta in electron subsystem of the crystal can be determined by
the equation $j^{\beta }(\xi ,t)=\left\langle \hat{\jmath}^{\beta }({\bf \xi
},t{\bf )}\right\rangle =Sp(\hat{\jmath}^{\beta }(\xi ,t)\hat{\rho}(t))$,
where $\beta =1,2,3$, $\hat{\rho}(t)$ is the density matrix of the crystal, $
\hat{\jmath}^{\beta }(\xi ,t)=\sum\limits_{s}\hat{\jmath}^{\beta }(\xi -\xi
_{s},t)$ is the $\beta $-th component of the current density operator in
electron subsystem (summation is carried out over all crystal electrons). In
the nonrelativistic approximation for the Dirac equation, the current
density operator of the electron $\hat{\jmath}^{\beta }(\xi -\xi _{s},t)$ \
is given by \cite{[8]}:
\begin{eqnarray}
&&\hat{\jmath}^{\beta }({\bf \xi }-{\bf \xi }_{s},t{\bf )=-}\frac{e^{2}}{mc}
\hat{n}({\bf \xi }-{\bf \xi }_{s})A^{\beta }({\bf \xi },t)+\hat{\jmath}
^{\beta }({\bf \xi }-{\bf \xi }_{s}),  \nonumber \\
&&\hat{\jmath}^{\beta }({\bf \xi }-{\bf \xi }_{s})=e[g^{\beta \mu }+\frac{i}{
2}g^{\beta k}g^{l\mu }\varepsilon _{kl\gamma }\sigma ^{\gamma }][\delta (
{\bf \xi }-{\bf \xi }_{s})\frac{\hat{p}_{\mu }}{2m}+\frac{\hat{p}_{\mu }}{2m}
\delta ({\bf \xi }-{\bf \xi }_{s})]  \label{19}
\end{eqnarray}
Here the value $-\frac{e^{2}}{mc}\hat{n}(\xi -\xi _{s})A^{\beta }(\xi ,t)$ \
is the potential (Rayleigh) part of the current operator, $\hat{n}(\xi -\xi
_{s})=\delta (\xi -\xi _{s})$ is the density operator of the $s$-th
electron,\ $\ A^{\beta }(\xi ,t)$ \ is the $\beta $-th component of the
electromagnetic field vector potential, $\hat{\jmath}^{\beta }(\xi -\xi
_{s}) $ is the sum of current operators of resonance electric and magnetic
transitions, $\varepsilon _{kl\gamma }$ is the asymmetric tensor ($
\varepsilon _{123}=1$), $\hat{p}_{\mu }=-i\hbar \nabla _{\mu }$ is the
momentum operator in the Lagrange co-ordinates system ($\nabla _{\mu }$ is
the covariant derivative), and $\sigma ^{\gamma },$ $\gamma =1,2,3$ are the
Pauli matrixes.

In the interaction representation, the von Neumann equation for the density
matrix of the crystal interacting with X-ray quanta has the form as follows:

\begin{eqnarray}
i\hbar \frac{\partial \hat{\rho}(t)}{\partial t} &=&[\hat{H}_{e\gamma
}^{\prime }(t),\hat{\rho}(t)],  \nonumber \\
\hat{H}_{e\gamma }^{\prime }(t) &=&e^{\frac{i}{\hbar }\hat{H}_{0}t}\hat{H}
_{e\gamma }(t)e^{-\frac{i}{\hbar }\hat{H}_{0}t}.  \label{20}
\end{eqnarray}
Here $\hat{H}_{0}$ stands for the unperturbed Hamiltonian of the crystal
while $\hat{H}_{e\gamma }(t)$ describes its interaction with X-ray quanta,

\[
\hat{H}_{e\gamma }(t)=-\frac{1}{c}\int d^{3}{\bf \xi }\sqrt{g({\bf \xi })}
\hat{\jmath}_{\beta }({\bf \xi },t{\bf )}A^{\beta }({\bf \xi },t{\bf ).}
\]
Accordingly to the perturbation theory the von Neumann equation solution can
be represented by the series $\hat{\rho}(t)=\sum\limits_{k=0}^{\infty }\hat{
\rho}^{(k)}(t)$, where $\hat{\rho}^{(k)}(t)\ $is proportional to the $k$-th
power of the electromagnetic field: $\ \hat{\rho}^{(k)}(t)$ $\thicksim (%
\hat{H}_{e\gamma }^{\prime }(t))^{k}\thicksim (E)^{k}$. Substituting this
series into Eq. (\ref{20}) and equating the terms of the same power in$\
\hat{H} _{e\gamma }^{\prime }(t)$, we obtain\ \
\[
i\hbar \frac{\partial \hat{\rho}^{(k)}(t)}{\partial t}=[\hat{H}_{e\gamma
}^{\prime }(t),\hat{\rho}^{(k-1)}(t)].
\]
Note that $\hat{H}_{e\gamma }^{\prime }(t)=0$ and ${\displaystyle{\frac{%
\partial \hat{\rho}(t) }{\partial t}}} =0$ in the absence of X-ray quanta,
i.e., $\hat{\rho}(t)=$ $\hat{\rho}_{0}$, where $\hat{\rho}_{0}$\ is the
unperturbed crystal density matrix. Further $\hat{\rho}_{0}$ is assumed to
be given by the equilibrium (Gibbs) density matrix, $\hat{\rho}_{0}=e^{\frac{%
F-\hat{H}_{0}}{kT}},$ where $F$ is the free energy and $T$ is the crystal
temperature. The average value of the current density operator in the
interaction representation can be obtained from the formula

\[
j^{\beta }(\xi ,t)=Sp(e^{\frac{i}{\hbar }\hat{H}_{0}t}\hat{\jmath}^{\beta
}(\xi ,t)e^{-\frac{i}{\hbar }\hat{H}_{0}t}\hat{\rho}(t))=Sp(\hat{\jmath}
_{H}^{\beta }(\xi ,t)\hat{\rho}_{0}),
\]
where $\hat{\jmath}_{H}^{\beta }({\bf \xi },t{\bf )}$ is the current
operator in the Heisenberg representation:

\begin{eqnarray}
&&\hat{\jmath}_{H}^{\beta }({\bf \xi },t{\bf )=}e^{\frac{i}{\hbar }\hat{H}
_{0}t}\hat{\jmath}^{\beta }({\bf \xi },t{\bf )}e^{-\frac{i}{\hbar }\hat{H}
_{0}t}  \nonumber \\
&&+\sum\limits_{n=0}^{\infty }\frac{1}{(i\hbar )^{n}}\int\limits_{-\infty
}^{t}dt_{1}\int\limits_{-\infty }^{t_{1}}dt_{2}\cdots \int\limits_{-\infty
}^{t_{n-1}}dt_{n}[\cdots \lbrack e^{\frac{i}{\hbar }\hat{H}_{0}t}\hat{\jmath}
^{\beta }({\bf \xi },t{\bf )}e^{-\frac{i}{\hbar }\hat{H}_{0}t},\hat{H}
_{e\gamma }^{\prime }(t_{1})],\cdots ,\hat{H}_{e\gamma }^{^{\prime
}}(t_{n})].  \nonumber
\end{eqnarray}
Here the time moments are put in the order: $t_{n}\leq t_{n-1}\leq \cdots
\leq t_{1}\leq t$.

In the linear approximation with respect to electromagnetic field, this
equation reduces to:
\begin{eqnarray}
&&j^{\beta }({\bf \xi },t)={\bf -}\frac{e^{2}}{mc}\sum\limits_{s}Sp(e^{\frac{
i}{\hbar }\hat{H}_{0}t}\hat{n}({\bf \xi }-{\bf \xi }_{s})e^{-\frac{i}{\hbar }
\hat{H}_{0}t}\hat{\rho}_{0})A^{\beta }({\bf \xi },t)  \nonumber \\
&&+\frac{i}{\hbar c}\sum\limits_{s}\sum\limits_{s^{^{\prime
}}}\int\limits_{-\infty }^{t}dt_{1}\int d^{3}{\bf \xi }^{\prime }\sqrt{g(
{\bf \xi }^{\prime })}\{Sp[e^{\frac{i}{\hbar }\hat{H}_{0}(t-t_{1})}\hat{
\jmath}^{\beta }({\bf \xi }-{\bf \xi }_{s})e^{-\frac{i}{\hbar }\hat{H}
_{0}(t-t_{1})}\hat{\jmath}_{\mu }({\bf \xi }^{\prime }-{\bf \xi }_{s^{\prime
}})\hat{\rho}_{0}]  \nonumber \\
&&-Sp\hat{\jmath}_{\mu }({\bf \xi }^{\prime }-{\bf \xi }_{s^{^{\prime }}})e^{%
\frac{i}{\hbar }\hat{H}_{0}(t-t_{1})}\hat{\jmath}^{\beta }({\bf \xi }-{\bf %
\xi }_{s})e^{-\frac{i}{\hbar }\hat{H}_{0}(t-t_{1})}\hat{\rho}_{0}]\}A^{\mu
}( {\bf \xi }^{\prime },t_{1}),  \nonumber
\end{eqnarray}
where the summation is carried out over all crystal electrons. Let us
present co-ordinates of the $s$-th electron in the following way: ${\bf \xi }
_{s}={\bf \xi }_{{\bf n}}+{\bf \xi }_{l}+{\bf \xi }_{k}+{\bf \eta }({\bf n}
,l)$ where ${\bf \xi }_{{\bf n}}$ stands for the co-ordinates of the unit
cell which the $s$-th electron belongs to, ${\bf \xi }_{l}$ are the
co-ordinates of the $l$-th atom inside this unit cell. In such situation, \
the identity $\sum\limits_{s}=\sum\limits_{{\bf n}}\sum\limits_{l}\sum
\limits_{k}$ holds true and, thus, the density of current induced by the
X-ray quanta in the crystal is represented by
\begin{eqnarray}
&&j^{\beta }({\bf \xi },t) ={\bf -}\frac{e^{2}}{mc}\sum\limits_{{\bf n}
}\sum\limits_{l}Sp(e^{\frac{i}{\hbar }\hat{H}_{0}t}\hat{n}_{l}({\bf \xi }-
{\bf \xi }_{{\bf n}}-{\bf \xi }_{l}-{\bf \eta }({\bf n},l))e^{-\frac{i}{
\hbar }\hat{H}_{0}t}\hat{\rho}_{0})A^{\beta }({\bf \xi },t)+\frac{i}{\hbar c}
\sum\limits_{{\bf n}}\sum\limits_{l}  \nonumber \\
&&\int\limits_{0}^{\infty }dt^{\prime }\int d^{3}{\bf \xi }^{\prime }\sqrt{%
g( {\bf \xi }^{\prime })}\{Sp[e^{\frac{i}{\hbar }\hat{H}_{0}t^{\prime }}\hat{
\jmath}_{l}^{\beta }({\bf \xi }-{\bf \xi }_{{\bf n}}-{\bf \xi }_{l}-{\bf %
\eta }({\bf n},l))e^{-\frac{i}{\hbar }\hat{H}_{0}t^{\prime }}\hat{\jmath}
_{l,\mu }({\bf \xi }^{\prime }-{\bf \xi }_{{\bf n}}-{\bf \xi }_{l}-{\bf \eta
}({\bf n},l))\hat{\rho}_{0}]  \nonumber \\
&&-Sp[\hat{\jmath}_{l,\mu }({\bf \xi }^{\prime }-{\bf \xi }_{{\bf n}}-{\bf %
\xi }_{l}-{\bf \eta }({\bf n},l))e^{\frac{i}{\hbar }\hat{H}_{0}t^{^{\prime
}}}\hat{\jmath}_{l}^{\beta }({\bf \xi }-{\bf \xi }_{{\bf n}}-{\bf \xi }_{l}-
{\bf \eta }({\bf n},l))e^{-\frac{i}{\hbar }\hat{H}_{0}t^{\prime }}\hat{\rho}
_{0}]\}A^{\mu }({\bf \xi }^{\prime },t-t^{\prime }).  \nonumber
\end{eqnarray}
Here $\hat{n}_{l}({\bf \xi }-{\bf \xi }_{{\bf n}}-{\bf \xi }_{l}-{\bf \eta }%
( {\bf n},l))=\sum\limits_{k}\hat{n}({\bf \xi }-{\bf \xi }_{{\bf n}}-{\bf %
\xi } _{l}-{\bf \xi }_{k}-{\bf \eta }({\bf n},l))$ is the density operator
of the electrons in $l$-th atom and $\hat{\jmath}_{l}^{\beta }({\bf \xi }-%
{\bf \xi } _{{\bf n}}-{\bf \xi }_{l}-{\bf \eta }({\bf n},l))=\sum\limits_{k}%
\hat{\jmath} ^{\beta }({\bf \xi }-{\bf \xi }_{{\bf n}}-{\bf \xi }_{l}-{\bf %
\xi }_{k}-{\bf \eta }({\bf n},l))$ is the current density operator in $l$-th
atom. In this equation we have neglected the components $Sp\left\{ [\hat{%
\jmath} _{l}^{\beta }({\bf n},t^{\prime }),\hat{\jmath}_{l^{\prime },\mu }(%
{\bf n} ^{\prime },0)]\hat{\rho}_{0}\right\} ,{\bf n\neq n}^{\prime }$ or $%
l\neq l^{\prime }$ describing the nonradiative migration of excitation from $%
l$-th atom to $l^{\prime }$-th one. In another words, we have neglected
delocalization of exciton in crystal. Besides, we leave out the
peculiarities of X-ray quanta interaction with the band electrons, assuming
all the electrons being localized on the atoms. These approximations are
proved to be correct for situation when the incident photon energy exceeds
significantly the energy of any resonance transition in the crystal. Indeed,
this condition allows us to restrict ourselves to the impulse approximation %
\cite{[9]} \ where there is no difference between energy spectra of band
electrons and electrons localized nearby the nuclei: the electrons density
periodicity in the crystal is of the only importance. \ That is why,
describing the interaction of X-ray quanta with the crystal, we use the
terminology appropriate rather to the isolated periodically arranged atoms
devoid of zone structure then to the crystal. It should be noted that the
above assumptions are not applicable to the X-ray spectroscopy where both
electron delocalization and peculiarities of the X-ray quanta interaction
with band electrons should be taken into consideration. However, this
problem is beyond the scope of this article and will be considered elsewhere.

\bigskip\ The remarks made above allow us to present the crystal Hamiltonian
$\hat{H}_{0}$ by $\hat{H}_{0}=\sum\limits_{{\bf n}}\sum\limits_{l}\hat{H}
_{l}+\hat{H}_{ph}$, where $\hat{H}_{ph}$ is the Hamiltonian of the phonon
subsystem and $\hat{H}_{l}$ is the Hamiltonian of the $l$-th atom in the
unit cell.

\bigskip For the atomic ground state in the crystal it is possible to
neglect the influence on this ground state of elementary crystal excitations
(including influence of interaction with phonons ) . In such a case, the
crystal density matrix $\hat{\rho}_{0}$ can be presented by the product $
\hat{\rho}_{0}=\hat{\rho}_{0}^{(e)}\hat{\rho}_{0}^{(ph)}$\ of electron
subsystem density matrix $\hat{\rho}_{0}^{(e)}=e^{\beta
(F^{(e)}-\sum\limits_{{\bf n}}\sum\limits_{l}\hat{H}_{l})}$ and phonon
subsystem density matrix $\hat{\rho}_{0}^{(ph)}=e^{\beta (F^{(ph)}-\hat{H}
_{ph})}$, where $\beta =1/kT$, and $F^{(e)},F^{(ph)}$ are the free electron
and phonon energies. As different from the ground state, the contribution of
crystal elementary excitations can not be neglected for excited atoms
because they lead to the appearance of the new decay channels of the exited
state and, consequently, increase the $\Gamma $-width of the exited state.
Therefore, we will assume the Hamiltonians of ground and exited states to be
different.

Now, let us discuss how the phonon spectrum is modified under effect of the
crystal deformation. In the harmonic approximation, the phonon subsystem
Hamiltonian of the deformed crystal in the Lagrange co-ordinate system takes
the form as follows:
\begin{equation}
H_{ph}=\sum\limits_{{\bf n}}\sum\limits_{l}\{g^{\alpha \beta }({\bf n},l)
\frac{p_{\alpha }({\bf n},l)p_{\beta }({\bf n},l)}{M_{l}}+\sum\limits_{{\bf %
n }^{\prime }}\sum\limits_{l^{\prime }}\Phi _{\alpha \beta }({\bf n},{\bf n}
^{\prime };l,l^{\prime })\eta ^{\alpha }({\bf n},l)\eta ^{\beta }({\bf n}
^{\prime },l^{\prime })\}  \label{21}
\end{equation}
where $p_{\alpha }({\bf n},l)$ is the $\alpha $-th component of the momentum
covector of the $l$ -th atom in the ${\bf n}$-th cell presented in the
Lagrange basis. Quantum-mechanically, the momentum is given by $p_{\alpha }(
{\bf n},l)=-i\hbar \nabla _{\alpha }$ with $\nabla _{\alpha }$ as the
covariant derivative. Further we assume the deformation tensor to be
constant over the unit cell; then, $g^{\alpha \beta }({\bf n},l)=\delta
^{\alpha \beta }+\varepsilon ^{\alpha \beta }({\bf n},l)\thickapprox
g^{\alpha \beta }({\bf n})$ . This approximation allows us to reduce the
Hamilton canonical equations to \
\[
\omega ^{2}B^{\alpha }({\bf n},l)=g^{\alpha \beta }({\bf n})\sum\limits_{%
{\bf n}^{\prime }}\sum\limits_{l^{\prime }}D_{\alpha \beta }({\bf n},{\bf n}
^{\prime };l,l^{\prime })B^{\beta }({\bf n}^{^{\prime }},l^{\prime }),
\]
where $B^{\alpha }({\bf n},l)=\sqrt{M_{l}}\eta ^{\alpha }({\bf n}
,l)e^{i\omega t}$ \ and $D_{\alpha \beta }({\bf n},{\bf n}^{\prime
};l,l^{\prime })= {\displaystyle{\frac{\Phi _{\alpha \beta }({\bf n},{\bf n}%
^{\prime };l,l^{\prime }) }{\sqrt{M_{l}M_{l^{\prime }}}}}} =D_{\alpha \beta
}({\bf n}-{\bf n}^{\prime };l,l^{\prime })$ is the dynamical matrix of the
crystal given in the Lagrange co-ordinates. \ The above equation for $%
B^{\alpha }({\bf n},l)$ can be rewritten as

\[
\omega ^{2}B^{\alpha }({\bf n},l)=(\delta ^{\alpha \beta }+\varepsilon
^{\alpha \beta })\sum\limits_{{\bf n}^{\prime }}\sum\limits_{l^{^{\prime
}}}D_{\alpha \beta }({\bf n}-{\bf n}^{\prime };l,l^{\prime })B^{\beta }({\bf %
n}^{\prime },l^{\prime }),
\]
where the deformation tensor $\varepsilon ^{\alpha \beta }$ is considered as
a parameter. Then the translation invariance of Hamiltonian (\ref{21})
allows us to write down $B^{\beta }({\bf n}^{\prime },l^{\prime };\hat{
\varepsilon})=e^{-ik_{i}\xi ^{i}({\bf n}-{\bf n}^{\prime })}B^{\alpha }({\bf %
n},l^{\prime };\hat{\varepsilon})$; this leads us to the equation

\[
\omega ^{2}B^{\alpha }({\bf n},l;\hat{\varepsilon})=(\delta ^{\alpha \gamma
}+\varepsilon ^{\alpha \gamma })\sum\limits_{l^{\prime }}D_{\gamma \beta }(
{\bf q};l,l^{\prime })B^{\beta }({\bf n},l^{\prime };\hat{\varepsilon}),
\]
where $D_{\gamma \beta }({\bf q};l,l^{\prime })=\sum\limits_{{\bf n}
}D_{\gamma \beta }({\bf n};l,l^{\prime })e^{-iq_{i}\xi ^{i}({\bf n})}$ is
the crystal lattice dynamical matrix in the Fourier representation. Hence,
we immediately obtain the dispersion equation

\begin{equation}
\det \left\| \omega ^{2}\delta ^{\alpha \beta }\delta _{l,l^{^{\prime
}}}-\delta ^{\alpha \gamma }\sum\limits_{l^{\prime }}D_{\gamma \beta }({\bf %
q };l,l^{\prime })-\varepsilon ^{\alpha \gamma }\sum\limits_{l^{\prime
}}D_{\gamma \beta }({\bf q};l,l^{^{\prime }})\right\| =0.  \label{22}
\end{equation}
As the Bravais lattice of the deformed crystal in the Lagrange basis
coincides with the Bravais lattice of the perfect crystal in the Euler
co-ordinates system, the matrix $\left\| D_{\gamma \beta }({\bf q}%
;l,l^{\prime })\right\| $ coincides with the dynamical matrix of the perfect
crystal. Consequently, dispersion equation (\ref{22}) does not contain
undefined quantities. Since the matrix $\left\| (\delta ^{\alpha \gamma
}+\varepsilon ^{\alpha \gamma })D_{\gamma \beta }({\bf q};l,l^{^{\prime
}})\right\| $ is Hermitian, the equation roots $\omega _{j}({\bf q},\hat{%
\varepsilon}),$ are real quantities; here $\ j=1,\cdots ,3r$ \ with $r$ as
the number of atoms in the crystal cell.

This matrix is Hermitian, its eigenvectors corresponding to these roots
satisfy the conditions of orthonormalization and fullness

\begin{eqnarray}
\sum\limits_{j}e^{\star \alpha }({\bf q},j;l;\hat{\varepsilon})e_{\beta }(
{\bf q},j;l^{\prime };\hat{\varepsilon}) &=&\delta _{\beta }^{\alpha }\delta
_{l,l^{\prime }}  \nonumber \\
\sum\limits_{l}e^{\star \alpha }({\bf q},j;l;\hat{\varepsilon})e_{\alpha }(
{\bf q},j^{\prime };l;\hat{\varepsilon}) &=&\delta _{j,j^{\prime }}
\nonumber
\end{eqnarray}
and allow us to implement the normal modes of the crystal $e^{\alpha }$ $(
{\bf q},j;{\bf n},l;\hat{\varepsilon})=\frac{1}{\sqrt{N}}e^{\alpha }$ $({\bf %
q},j;l;\hat{\varepsilon})e^{ik_{i}\xi ^{i}({\bf n})}$. For these modes we
have

\begin{eqnarray}
\sum\limits_{{\bf q,}j}e^{\star \alpha }({\bf q},j;{\bf n},l;\hat{%
\varepsilon })e_{\beta }({\bf q},j;{\bf n}^{\prime },l^{\prime };\hat{%
\varepsilon}) &=&\delta _{\beta }^{\alpha }\delta _{l,l^{\prime }}\delta _{%
{\bf n},{\bf n} ^{\prime }}  \nonumber \\
\sum\limits_{{\bf n},l}e^{\star \alpha }({\bf q},j;{\bf n},l;\hat{%
\varepsilon })e_{\alpha }({\bf q}^{\prime },j^{\prime };{\bf n},l;\hat{%
\varepsilon}) &=&\delta _{j,j^{\prime }}\Delta ({\bf q}-{\bf q}^{\prime })
\nonumber
\end{eqnarray}
where $\Delta ({\bf q})=\sum\limits_{{\bf l}}\delta _{{\bf q},{\bf \tau }(
{\bf l})}$ and ${\bf \tau }({\bf l})=( {\displaystyle{\frac{2\pi l_{1} }{%
a_{1}}}} , {\displaystyle{\frac{2\pi l_{2} }{a_{2}}}} , {\displaystyle{\frac{%
2\pi l_{3} }{a_{3}}}} )$ is the vector of the crystal reciprocal lattice.
Next, using the standard method we obtain the Hamiltonian of the crystal
phonon subsystem in the harmonic approximation

\begin{equation}
\hat{H}_{ph}(\varepsilon )=\sum\limits_{{\bf q,}j}\hbar \omega _{j}({\bf q},
\hat{\varepsilon})[b_{{\bf q,}j}^{\dagger }(\hat{\varepsilon})b_{{\bf q,}j}(
\hat{\varepsilon})+\frac{1}{2}]  \label{23}
\end{equation}
where $b_{{\bf q,}j}^{\dagger }(\hat{\varepsilon})$ and $b_{{\bf q,}j}(\hat{
\varepsilon})$ are the phonon creation and annihilation operators. It is
easy to verify that the proposed algorithm of accounting the influence of
deformation on the crystal phonon spectrum also suits the case of anharmonic
oscillations of the crystal lattice. At the end we will make two remarks.
The first: since the deformation tensor $\varepsilon ^{\alpha \beta
}=\varepsilon ^{\alpha \beta }({\bf \xi })$ is the essence of the smooth
co-ordinates function, the equations (\ref{22})-(\ref{23}) are local. The
discussed above method of accounting of deformation influence on the crystal
phonon spectrum is the approximation, that can be considered precise enough
only when the length of the phonon free path in the crystal is less then the
linear dimensions of the area where the deformation tensor variation can be
observed. The second remark concerns the fact that the introduced parametric
method of accounting the deformation influence on the crystal phonon
spectrum is not something especially new in the solid-state physics: this
method is being the standard one in accounting the deformation influence on
the energy spectrum of band electrons (see for example \cite{[10]}).

After the discussion made above let us continue the calculations of the
current in the deformed crystals that was induced by X-ray quanta. First let
us take up the Rayleigh component of the current

\[
j_{Rl}^{\beta }({\bf \xi },t)={\bf -}\frac{e^{2}}{mc}\sum\limits_{{\bf n}
}\sum\limits_{l}Sp(e^{\frac{i}{\hbar }\hat{H}_{0}t}\hat{n}_{l}({\bf \xi }-
{\bf \xi }_{{\bf n}}-{\bf \xi }_{l}-{\bf \eta }({\bf n},l))e^{-\frac{i}{
\hbar }\hat{H}_{0}t}\hat{\rho}_{0})A^{\beta }({\bf \xi },t),
\]
that makes the main contribution. As $e^{-\frac{i}{\hbar }\hat{H}_{0}t}\hat{
\rho}_{0}e^{\frac{i}{\hbar }\hat{H}_{0}t}=\hat{\rho}_{0}$ and ${\bf E}({\bf %
\xi },\omega )= {\displaystyle{\frac{i\omega }{c}}} {\bf A}({\bf \xi }%
,\omega )$ in the Coulomb calibration, using the Fourier series $\hat{n}_{l}(%
{\bf \xi }-{\bf \xi }_{{\bf n}}-{\bf \xi }_{l}-{\bf \eta }({\bf n},l))=\int
d^{3}{\bf k}\hat{n}_{l}({\bf k})e^{ik_{i}\xi ^{i}}e^{-ik_{i}\xi ^{i}({\bf n}%
)}e^{-ik_{i}\xi _{l}^{i}}e^{-ik_{i}\eta ^{i}( {\bf n},l)}$, we obtain the
formula for Rayleigh component of the current
\[
j_{Rl}^{\beta }({\bf \xi },\omega )= {\displaystyle{\frac{ie^{2} }{m\omega }}%
} \int d^{3}{\bf k}e^{ik_{i}\xi ^{i}}\sum\limits_{{\bf n}}e^{-ik_{i}\xi
^{i}( {\bf n})}\sum\limits_{l}e^{-ik_{i}\xi _{l}^{i}}Sp(\hat{n}_{l}({\bf k})%
\hat{ \rho}_{0}^{(e)})Sp(e^{-ik_{i}\eta ^{i}({\bf n},l)}\hat{\rho}
_{0}^{(ph)})E^{\beta }({\bf \xi },\omega ).
\]
The value $Sp(e^{-ik_{i}\eta ^{i}({\bf n},l)}\hat{\rho}
_{0}^{(ph)})=e^{-W_{l}({\bf k},{\bf \xi })}$ is the essence of Debye-Waller
factor where $W_{l}({\bf k},{\bf \xi })$ is to be determined by the
equation\ \ \ \ \ \ \ \ \ \ \ \ \ \ \ \ \ \ \ \ \ \ \ \ \
\begin{equation}
W_{l}({\bf k},{\bf \xi })= {\displaystyle{\frac{\hbar }{2M_{l}N}}}
\sum\limits_{{\bf q},j}\frac{\left| k_{\alpha }e^{\alpha }({\bf q},j;l;\hat{
\varepsilon})\right| ^{2}}{\omega _{j}({\bf q},\hat{\varepsilon})}\coth {%
\displaystyle{\frac{\hbar \omega _{j}({\bf q},\hat{\varepsilon}) }{2k_{B}T}}}
\label{24}
\end{equation}
Accordingly the Debye-Waller factor in the deformed crystals is the function
of the deformation tensor and consequently is the function of co-ordinates
determined by the equations (\ref{22})-(\ref{23}). As $Sp(\hat{n}_{l}({\bf k}
)\hat{\rho}_{0}^{(e)})=f_{l}({\bf k})$ is the form factor of the $l$-th atom
unit cell, the standard formula $\sum\limits_{{\bf n}}e^{-ik_{i}\xi ^{i}(
{\bf n})}=N_{cell}\sum\limits_{{\bf l}}\delta ({\bf k}-{\bf \tau }^{({\bf l}
)})$, where $N_{cell}$ is the number of the unit cells in the volume unit,
allows us to obtain the final formula for the Rayleigh component of the
current

\[
j_{Rl}^{\beta }({\bf \xi },\omega )= {\displaystyle{\frac{ie^{2}N_{cell} }{%
m\omega }}} \sum\limits_{{\bf l}}(\sum\limits_{l}e^{-W_{l}({\bf \tau }^{(%
{\bf l})},{\bf \xi })}f_{l}({\bf \tau }^{({\bf l})})e^{-i\tau _{i}^{({\bf l}%
)}\xi _{l}^{i}})e^{-i\tau _{i}^{({\bf l})}\xi ^{i}}E^{\beta }({\bf \xi }%
,\omega )
\]
or the formula for the Rayleigh component of the deformed crystal
polarization tensor in the X-ray frequency range

\begin{equation}
\chi _{Rl}({\bf \xi },{\bf \xi }^{\prime },\omega )=- {\displaystyle{\frac{%
4\pi e^{2}N_{cell} }{m\omega ^{2}}}} \sum\limits_{{\bf l}}(\sum%
\limits_{l}e^{-W_{l}({\bf \tau }^{({\bf l})},{\bf \xi })}f_{l}({\bf \tau }^{(%
{\bf l})})e^{-i\tau _{i}^{({\bf l})}\xi _{l}^{i}})e^{-i\tau _{i}^{({\bf l}%
)}\xi ^{i}}\delta ({\bf \xi }-{\bf \xi } ^{\prime })  \label{25}
\end{equation}

In the frequency range where the X-ray quanta energy is greater then the
binding energy of any electron in the crystal, the current component
describing the absorption of quanta by the atoms and the consequent decay of
the exited state which atom passed into by means of quantum absorption is
usually much smaller then the Rayleigh component (in the X-ray optics it is
called the dispersion correction). Nevertheless the accounting of this
component is very important as it determines the absorption of X-ray quanta
in this crystal. Using the Fourier decomposition $\hat{\jmath}_{l}^{\beta }(
{\bf \xi }-{\bf \xi }_{{\bf n}}-{\bf \xi }_{l}-{\bf \eta }({\bf n},l))=\int
d^{3}{\bf k}\hat{\jmath}_{l}^{\beta }({\bf k})e^{ik_{i}\xi
^{i}}e^{-ik_{i}\xi ^{i}({\bf n})}e^{-ik_{i}\xi _{l}^{i}}e^{-ik_{i}\eta ^{i}(
{\bf n},l)}$, this current component can be written down in the following way

\begin{eqnarray}
&&j_{disp}^{\beta }({\bf \xi },\omega )=\frac{1}{\hbar \omega }\int d^{3}
{\bf \xi }^{\prime }\sqrt{g({\bf \xi }^{\prime })}\sum\limits_{{\bf n}
}\sum\limits_{l}\int\limits_{0}^{\infty }dt^{\prime }\int d^{3}{\bf k}\int
d^{3}{\bf k}^{\prime }e^{ik_{i}\xi ^{i}}e^{ik_{i}^{\prime }\xi ^{^{\prime
}i}}e^{-i(k_{i}+k_{i}^{\prime })\xi ^{i}({\bf n})^{\prime
}}e^{-i(k_{i}+k_{i}^{\prime })\xi _{l}^{i}}  \nonumber \\
&&\times \{Sp[e^{\frac{i}{\hbar }(\hat{H}_{l}+\hat{H}_{ph}({\bf \xi )}
)t^{^{\prime }}}e^{-ik_{i}\eta ^{i}({\bf n},l)}\hat{\jmath}_{l}^{\beta }(
{\bf k})e^{-\frac{i}{\hbar }(\hat{H}_{l}+\hat{H}_{ph}({\bf \xi }))t^{\prime
}}\hat{\jmath}_{l,\mu }({\bf k}^{\prime })e^{-ik_{i}^{\prime }\eta ^{i}({\bf %
n},l)}\rho _{0}^{(e)}(l)\hat{\rho}_{0}^{(ph)}({\bf \xi })]  \nonumber \\
&&-Sp[\hat{\jmath}_{l,\mu }({\bf k}^{\prime })e^{-ik_{i}^{\prime }\eta ^{i}(
{\bf n},l)}e^{\frac{i}{\hbar }(\hat{H}_{l}+\hat{H}_{ph}({\bf \xi )}
)t^{\prime }}e^{-ik_{i}\eta ^{i}({\bf n},l)}\hat{\jmath}_{l}^{\beta }({\bf k}%
)e^{-\frac{i}{\hbar }(\hat{H}_{l}+\hat{H}_{ph}({\bf \xi }))t^{\prime }}\rho
_{0}^{(e)}(l)\hat{\rho}_{0}^{(ph)}({\bf \xi })]\}e^{i\omega t^{\prime
}}E^{\mu }({\bf \xi }^{\prime },\omega )  \nonumber
\end{eqnarray}
In the X-ray frequency range under the consideration $\Gamma $ (the width of
the exited atom state in the crystal) is about $10^{16}\frac{1}{\sec }$,
that is why in the X-ray diffraction optics the assumption $\hat{H}_{l}+%
\hat{H}_{ph}({\bf \xi )\thickapprox }$ $\hat{H}_{l}$ is being considered.
This assumption means that we neglect the influence of Raman scattering of
X-ray quanta on the phonons. If we take this assumption and use the standard
formulas $\sum\limits_{{\bf n}}e^{-i(k_{i}+k_{i}^{\prime })\xi ^{i}({\bf n}
)}=N_{cell}\sum\limits_{{\bf l}}\delta ({\bf k}+{\bf k}^{\prime }-{\bf \tau }
^{({\bf l})})$ and $Sp(e^{-i(k_{i}+k_{i}^{\prime })\eta ^{i}({\bf n},l)}\hat{
\rho}_{0}^{(ph)}({\bf \xi }))=e^{-W_{l}({\bf k}+{\bf k}^{^{\prime }},{\bf %
\xi })}$, we obtain for $j_{disp}^{\beta }({\bf \xi },\omega )$ the
following equation

\begin{eqnarray}
&&j_{disp}^{\beta }({\bf \xi },\omega )=  \nonumber \\
&&\frac{N_{cell}}{\hbar \omega }\sum\limits_{{\bf l}}e^{-W_{l}({\bf \tau }%
^{( {\bf l})},{\bf \xi })}\sum\limits_{l}e^{-i\tau _{i}^{({\bf l})}\xi
_{l}^{i}}\int d^{3}{\bf \xi }^{\prime }\sqrt{g({\bf \xi }^{\prime })}
e^{-i\tau _{i}^{({\bf l})}\xi ^{^{\prime }i}}\int d^{3}{\bf k}e^{ik_{i}(\xi
^{i}-\xi ^{^{\prime }i})}\frac{1}{2J_{a}(l)+1}  \nonumber \\
&&\sum\limits_{M_{a}=-J_{a}(l)}^{J_{a}(l)}\sum\limits_{\lambda (a)}\rho
(\lambda (a))[\left\langle a(l);J_{a}(l),M_{a}\right| \hat{\jmath}
_{l}^{\beta }({\bf k})\int\limits_{0}^{\infty }dt^{\prime }e^{-\frac{i}{
\hbar }(\hat{H}_{l}-E_{a}-\hbar \omega )t^{\prime }}\hat{\jmath}_{l,\mu }(-
{\bf k}-{\bf \tau }^{({\bf l})})\left| a(l);J_{a}(l),M_{a}\right\rangle
\nonumber \\
&&-\left\langle a(l);J_{a}(l),M_{a}\right| \hat{\jmath}_{l,\mu }(-{\bf k}-
{\bf \tau }^{({\bf l})})\int\limits_{0}^{\infty }dt^{\prime }e^{\frac{i}{
\hbar }(\hat{H}_{l}-E_{a}+\hbar \omega )t^{\prime }}\hat{\jmath}_{l}^{\beta
}({\bf k})\left| a(l);J_{a}(l),M_{a}\right\rangle ]E^{\mu }({\bf \xi }
^{\prime },\omega )  \nonumber
\end{eqnarray}
where $a(l);J_{a}(l),M_{a}$ is the $l$-th atom ground state vector, $\rho
(\lambda (a))$ is the static weight of the ground state with the full set of
quantum numbers $\lambda (a)$ except for the angular momentum, $E_{a}$ is
the energy of this state, $J_{a}(l)$ is the total angular momentum of the
atom, $M_{a}$ is the projection of the total angular momentum on the
quantization axis.

Having used the formalism developed in \cite{[9]} for the description of
decay and the lifetime of virtual states of quantum systems in the case,
when the spectrum of the Hamiltonian of the system is continuous, we obtain
the following formula
\begin{eqnarray}
&&\left\langle a(l);J_{a}(l),M_{a}\right| \hat{\jmath}_{l}^{\beta }({\bf k}
)\int\limits_{0}^{\infty }dt^{\prime }e^{-\frac{i}{\hbar }(\hat{H}
_{l}-E_{a}-\hbar \omega )t^{\prime }}\hat{\jmath}_{l,\mu }(-{\bf k}-{\bf %
\tau }^{({\bf l})})\left| a(l);J_{a}(l),M_{a}\right\rangle  \nonumber \\
&&-\left\langle a(l);J_{a}(l),M_{a}\right| \hat{\jmath}_{l,\mu }(-{\bf k}-
{\bf \tau }^{({\bf l})})\int\limits_{0}^{\infty }dt^{\prime }e^{\frac{i}{
\hbar }(\hat{H}_{l}-E_{a}+\hbar \omega )t^{\prime }}\hat{\jmath}_{l}^{\beta
}({\bf k})\left| a(l);J_{a}(l),M_{a}\right\rangle =\frac{\hbar }{2i}
\sum\limits_{J_{b},M_{b}}  \nonumber \\
&&\int \rho (E_{b})dE_{b}\left\langle a(l);J_{a}(l),M_{a}\right| \hat{\jmath}
_{l}^{\beta }({\bf k})\left| b(l);J_{b}(l),M_{b}\right\rangle \left\langle
a(l);J_{a}(l),M_{a}\right| \hat{\jmath}_{l,\mu }({\bf k}+{\bf \tau }^{({\bf %
l })}\left| b(l);J_{b}(l),M_{b}\right\rangle ^{\ast }  \nonumber \\
&&\times \frac{1}{E_{a}+\hbar \omega -E_{b}+R_{b}^{(+)}(E_{a}+\hbar \omega )}
+\frac{\hbar }{2i}\sum\limits_{J_{b},M_{b};J_{c},M_{c}}\int \rho
(E_{b})dE_{b}\int \rho (E_{c})\delta (E_{c}-\hbar \omega -E_{a})dE_{c}
\nonumber \\
&&\times \left\langle a(l);J_{a}(l),M_{a}\right| \hat{\jmath}_{l}^{\beta }(
{\bf k})\left| c(l);J_{c}(l),M_{c}\right\rangle \left\langle
a(l);J_{a}(l),M_{a}\right| \hat{\jmath}_{l,\mu }({\bf k}+{\bf \tau }^{({\bf %
l })}\left| b(l);J_{b}(l),M_{b}\right\rangle ^{\ast }  \nonumber \\
&&\times R_{cb}^{(+)}(E_{c})\frac{1}{E_{a}+\hbar \omega
-E_{b}+R_{b}^{(+)}(E_{a}+\hbar \omega )}  \label{26}
\end{eqnarray}
Here $\rho (E_{b})$ and $\rho (E_{c})$ are the densities of the exited atom
states with energies $E_{b}$ and $E_{c}$ accordingly, $R_{b}^{(+)}(E)=\lim_{
\epsilon \rightarrow +0}\left\langle b(l);J_{b}(l),M_{b}\right| \hat{R}
(E+i\epsilon )\left| b(l);J_{b}(l),M_{b}\right\rangle =D_{b}(E)-iI_{b}(E)$
is the matrix diagonal element of the level shift operator determined by the
equation \cite{[9]}

\[
\hat{R}(E)=\hat{V}_{int}+\hat{V}_{int}\frac{1-\hat{P}_{b}}{E-\hat{H}_{l}}
\hat{R}(E),
\]
where $\hat{V}_{int}$ is the Hamiltonian of the interaction of the $l$-th
atom in the exited state with the crystal, $\hat{P}_{b}=\left|
b(l);J_{b}(l),M_{b}\right\rangle \left\langle b(l);J_{b}(l),M_{b}\right| $
is the operator of the projection on the state $b(l);J_{b}(l),M_{b}$. The
imaginary part of matrix element $R_{b}^{(+)}(E)$ is determined by formula
\[
\mathop{\rm Im} R_{b}^{(+)}(E)=I_{b}(E)=\frac{\hbar \Gamma }{2}=\pi
\sum\limits_{c}\int \rho (E_{c})\left| \left\langle
c(l);J_{c}(l),M_{c}\right| \hat{R} ^{(+)}(E_{c})\left|
b(l);J_{b}(l),M_{b}\right\rangle \right| ^{2},
\]
where the summation is taken over all decay channels of the excited atom
state and is the essence of $\Gamma $ (the width of the crystal atom excited
state). The term in the first part of formula (\ref{26}) is proportional to $
R_{cb}^{(+)}(E_{a}+\hbar \omega )$ and takes into account the interaction of
the atom in the excited state with the crystal and in particular allows to
account such an effect as EXAFS in the crystal polarization tensor. To
describe this interaction it is enough to let $R_{cb}^{(+)}(E_{c})=\left
\langle c(l);J_{c}(l),M_{c}\right| {\cal \hat{V}}\left|
b(l);J_{b}(l),M_{b}\right\rangle $, where ${\cal \hat{V}}$ is the Watson
pseudo-potential determined by the equation \cite{[9]}:

\[
{\cal \hat{V}}=\sum\limits_{l^{\prime }\neq l}\left\langle 0\right| \hat{T}
_{l^{\prime }}\left| 0\right\rangle +\sum\limits_{l^{^{\prime \prime }}\neq
l^{\prime }\neq l}\left\langle 0\right| \hat{T}_{l^{\prime }}\frac{1-\hat{P}
_{0}}{\hat{d}}\hat{T}_{l^{^{\prime \prime }}}\left| 0\right\rangle +\cdots ,
\]
where $\hat{T}_{l^{\prime }}$ is the matrix of scattering by $l^{\prime }$
-th crystal atom, $\left| 0\right\rangle $ is the crystal ground state
vector, $\hat{P}_{0}=\left| 0\right\rangle \left\langle 0\right| $ is the
operator of the projection on the crystal ground state, $\frac{1}{\hat{d}}$
is the propagator (for details look at \cite{[9]}). Here we will restrict
ourself only with the first term of (\ref{26})-th right part since the
description of the contribution of interaction of exited atom with the
crystal to the crystal polarization tensor is worth writing a separate
article.

For the further calculations we will use the atom current density operator
multipole decomposition \cite{[11]},\cite{[12]}

\begin{eqnarray}
&&\hat{\jmath}_{l}^{\beta }({\bf k})=4\pi \sum\limits_{L,M}(-i)^{L}\{ {%
\displaystyle{\frac{ck^{L}(L+1) }{(2L+1)!!\sqrt{L(L+1)}}}} \hat{E}%
_{L,M}(k)F_{L,M}^{(e)\beta }({\bf \bar{k}})+  \nonumber \\
&&i {\displaystyle{\frac{ck^{L}(L+1) }{(2L+1)!!\sqrt{L(L+1)}}}} \hat{M}%
_{L,M}(k)F_{L,M}^{(m)\beta }({\bf \bar{k}})+ {\displaystyle{\frac{k^{L} }{%
(2L+1)!!}}} \hat{Q}_{L,M}(k)F_{L,M}^{(ch)\beta }({\bf \bar{k}})\}  \nonumber
\end{eqnarray}
where $L$ is the multipolarity, $-L$ $\leqslant M\leqslant L$; $\hat{E}
_{L,M}(k),\hat{M}_{L,M}(k)$ and $\hat{Q}_{L,M}(k)$ are the operators of the
electric, magnetic and charge momenta of a multipole atom; ${\bf F}
_{L,M}^{(e,m,ch)}({\bf \bar{k}})$ are the spherical vectors of electric,
magnetic and charge type \cite{[6]},\cite{[9]}, ${\bf \bar{k}=}\frac{{\bf k}
}{k}$ is the unit covector in the covector ${\bf k}$ direction. Since the
field of X-ray quanta in the crystal with a high precision can be treated as
transversal one the components with $\hat{Q}_{L,M}(k)$ in this decomposition
can be neglected right away. As the magnet multipole transitions in atoms
are much smaller then the electric multipole transitions we will content our
self only with the electric multipoles. Having used the spherical vectors $%
{\bf F}_{L,M}^{(e)}({\bf \bar{k}})$ properties, Wigner-Eckart and Ziggert
theorems, we obtain the equation

\begin{eqnarray}
&&\frac{1}{2J_{a}+1}\sum\limits_{M_{a},M_{b}}\left\langle
a(l);J_{a}(l),M_{a}\right| \hat{\jmath}_{l}^{\beta }({\bf k})\left|
b(l);J_{b}(l),M_{b}\right\rangle  \nonumber \\
&&\times \left\langle a(l);J_{a}(l),M_{a}\right| \hat{\jmath}_{l,\mu }({\bf %
k }+{\bf \tau }^{({\bf l})}\left| b(l);J_{b}(l),M_{b}\right\rangle ^{\ast }
\nonumber \\
&=&\frac{2\pi }{(2J_{a}+1)}\sum\limits_{L}\left( {\displaystyle{\frac{(L+1)
}{(2L+1)!!\sqrt{L(L+1)}}}} \right) ^{2}c^{2}k^{L}k({\bf l})^{L}\left\langle
a(l);J_{a}(l)\right| \hat{Q} _{L}(k)\left| b(l);J_{b}(l)\right\rangle
\nonumber \\
&&\times \left\langle a(l);J_{a}(l)\right| \hat{Q}_{L}(k({\bf l}))\left|
b(l);J_{b}(l)\right\rangle ^{\ast }\Pi _{\mu }^{\beta }({\bf \bar{k},\bar{k}}
({\bf l});L)  \label{27}
\end{eqnarray}
where $\left\langle a(l);J_{a}(l)\right| \hat{Q}_{L}(k)\left|
b(l);J_{b}(l)\right\rangle $ and $\left\langle a(l);J_{a}(l)\right| \hat{Q}
_{L}(k({\bf l}))\left| b(l);J_{b}(l)\right\rangle $ are the reduced matrix
elements of the multipole charge momenta of the crystal unit cell $l$-th
atom; ${\bf \bar{k}}({\bf l})=\frac{{\bf k}+{\bf \tau }^{({\bf l})}}{k({\bf %
l })}$ is the unit covector along covector $({\bf k}+{\bf \tau }^{({\bf l}%
)}) $ , $k=\sqrt{g^{\alpha \beta }k_{\alpha }k_{\beta }},k({\bf l})=\sqrt{
g^{\alpha \beta }(k_{\alpha }+\tau _{\alpha }^{({\bf l})})(k_{\beta }+\tau
_{\beta }^{({\bf l})})}$ are these covectors' lengths; $\left\| \Pi _{\mu
}^{\beta }({\bf \bar{k},\bar{k}}({\bf l});L)\right\| $ is the polarization
matrix with its elements in spiral bases ${\bf e}^{\pm 1}=\frac{1}{\sqrt{2}}%
( {\bf n}_{1}\pm {\bf n}_{2}),{\bf e}^{\pm 1}({\bf l})=\frac{1}{\sqrt{2}}(%
{\bf n}_{1}({\bf l})\pm {\bf n}_{2}({\bf l}))$, where ${\bf n}_{1,2}$ and $%
{\bf n} _{1,2}({\bf l})$ are the orthonormalized covectors in the planes
orthogonal to ${\bf \bar{k}}$ and ${\bf \bar{k}}({\bf l})$ accordingly are
expressed through the Wigner functions

\[
\Pi _{\mu }^{\beta }({\bf \bar{k},\bar{k}}({\bf l});L)=\sum
\limits_{M=-L}^{L}D_{\mu ,M}^{L}({\bf \bar{k}}({\bf l}))D_{\beta ,M}^{L\ast
}({\bf \bar{k}}).
\]
If to choose the axis $z$ in the covectors ${\bf k}$ and ${\bf k}+{\bf \tau }
({\bf l})$ plane and to mark these covectors' angles relative to the axis $z$
as $\vartheta $ and $\vartheta _{1\text{,}}$ the formulas for the elements
of the polarization matrix are sufficiently simplified:

\begin{equation}
\Pi _{\mu }^{\beta }({\bf \bar{k},\bar{k}}({\bf l});L)=\sum
\limits_{M=-L}^{L}D_{\mu ,M}^{L}({\bf \bar{k}}({\bf l}))D_{\beta ,M}^{L\ast
}({\bf \bar{k}})=d_{\beta ,\mu }^{L}(\vartheta -\vartheta _{1})  \label{27a}
\end{equation}
By substituting the formulas (\ref{26})-(\ref{27a}) to the expression for $
j_{disp}^{\beta }({\bf \xi },\omega )$, we obtain

\begin{eqnarray}
&&j_{disp}^{\beta }({\bf \xi },\omega )=\frac{N_{cell}}{i\omega }
\sum\limits_{{\bf l}}e^{-W_{l}({\bf \tau }^{({\bf l})},{\bf \xi }
)}\sum\limits_{l}e^{-i\tau _{i}^{({\bf l})}\xi _{l}^{i}}\int d^{3}{\bf \xi }
^{\prime }\sqrt{g({\bf \xi }^{\prime })}e^{-i\tau _{i}^{({\bf l})}\xi
^{^{\prime }i}}\int d^{3}{\bf k}e^{ik_{i}(\xi ^{i}-\xi ^{^{\prime }i})}
\nonumber \\
&&\times \frac{2\pi }{2J_{a}(l)+1}\sum\limits_{\lambda (a)}\rho (\lambda
(a))\sum\limits_{L}\left( {\displaystyle{\frac{(L+1) }{(2L+1)!!\sqrt{L(L+1)}}%
}} \right) ^{2}\times  \nonumber \\
&&\sum\limits_{J_{b}=\left| L-J_{a}\right| }^{\left| L+J_{a}\right| }\int
\rho (E_{b})dE_{b}\frac{c^{2}k^{L}k({\bf l})^{L}\left\langle
a(l);J_{a}(l)\right| \hat{Q}_{L}(k)\left| b(l);J_{b}(l)\right\rangle
\left\langle a(l);J_{a}(l)\right| \hat{Q}_{L}(k({\bf l}))\left|
b(l);J_{b}(l)\right\rangle ^{\ast }}{E_{a}+\hbar \omega -E_{b}+D(E_{a}+\hbar
\omega )-i\frac{\hbar }{2}\Gamma (E_{a}+\hbar \omega )}  \nonumber \\
&&\times d_{\beta ,\mu }^{L}(\vartheta -\vartheta _{1})E^{\mu }({\bf \xi }
^{\prime },\omega )  \label{28}
\end{eqnarray}
Then the polarization tensor of the deformed crystal in the X-ray frequency
range is determined by the equation

\begin{eqnarray}
&&\chi _{\mu }^{\beta }({\bf \xi },{\bf \xi }^{\prime },\omega
)=\sum\limits_{{\bf l}}\sum\limits_{l}e^{-W_{l}({\bf \tau }^{({\bf l})},{\bf %
\xi })}e^{-i\tau _{i}^{({\bf l})}\xi _{l}^{i}}e^{-i\tau _{i}^{({\bf l})}\xi
^{^{\prime }i}}\{- {\displaystyle{\frac{4\pi e^{2}N_{cell} }{m\omega ^{2}}}}
f_{l}({\bf \tau }^{({\bf l})})\delta ({\bf \xi }-{\bf \xi }^{^{\prime
}})\delta _{\mu }^{\beta }  \nonumber \\
&&+\frac{4\pi N_{cell}}{\omega ^{2}}\frac{2\pi }{2J_{a}(l)+1}
\sum\limits_{\lambda (a)}\rho (\lambda (a))\sum\limits_{L}\left( {%
\displaystyle{\frac{(L+1) }{(2L+1)!!\sqrt{L(L+1)}}}} \right)
^{2}\sum\limits_{J_{b}=\left| L-J_{a}\right| }^{\left| L+J_{a}\right| }\int
\rho (E_{b})dE_{b}  \nonumber \\
&&\times \int d^{3}{\bf k}e^{ik_{i}(\xi ^{i}-\xi ^{^{\prime }i})}\frac{
c^{2}k^{L}k({\bf l})^{L}\left\langle a(l);J_{a}(l)\right| \hat{Q}
_{L}(k)\left| b(l);J_{b}(l)\right\rangle \left\langle a(l);J_{a}(l)\right|
\hat{Q}_{L}(k({\bf l}))\left| b(l);J_{b}(l)\right\rangle ^{\ast }}{
E_{a}+\hbar \omega -E_{b}+D(E_{a}+\hbar \omega )-i\frac{\hbar }{2}\Gamma
(E_{a}+\hbar \omega )}  \nonumber \\
&&\times d_{\beta ,\mu }^{L}(\vartheta -\vartheta _{1})\}  \label{29}
\end{eqnarray}

Let us discuss the obtained expression. First of all let us note that in ( %
\ref{29}) the covector ${\bf k}$ and the frequency $\omega $ are not bound
together by any equation. As it is seen from (\ref{25}), (\ref{28}) and (\ref%
{29}), the crystal polarizability in the X-ray frequency range is a scalar
value and the relation between the current induced in the crystal by X-ray
quanta and electric vector or between polarization vector and electric
vector is local only if we neglect the so-called dispersion correction and,
consequently, we neglect the crystal absorption of X-ray quanta. If we want
to account the absorption in the crystal, the statements like ''the
polarization vector in the X-ray frequency range is related to the electric
vector by the equation ${\bf P}({\bf r},\omega )=\chi ({\bf r},\omega ){\bf %
E }({\bf r},\omega )$, where $\chi ({\bf r},\omega )$ is the crystal
polarizability, and is being a scalar periodic function'' are just
incorrect. When we take into account the dispersion correction, we must
understand that the crystal polarizability in the X-ray frequency range is
in principle the tensor but not the scalar value and the relation between
the polarization vector and the electric vector is unlocal. That is why the
procedure of derivation of the equations (\ref{6}) and (\ref{15}) in parts $
1 $ and $2$ (they describe the dynamic diffraction of the electromagnetic
fields on the media with perfect periodic and quasi-periodic structures)
accounts the nonlocality of the response of the medium to the
electromagnetic field and is also true for the X-ray quanta diffraction on
the deformed crystals. The polarizability tensor of the medium is not being
a member of the equations (\ref{6}) and (\ref{15}) but his
Fourier-components $\hat{\chi}_{{\bf l}}({\bf k},\omega )$, that have the
relation between the wave covector ${\bf k}$ and the frequency $\omega $\
already defined: $k^{2}=k_{i}k^{i}=\frac{\omega ^{2}}{c^{2}}$. If to note
that electromagnetic fields with wave covectors ${\bf k}$ and ${\bf k+\tau }
^{{\bf (l)}}$ interact effectively only in vicinity to the Bragg conditions $
k^{2}=k({\bf l})^{2}=(k_{i}+\tau _{i}^{{\bf (l)}}{\bf )(}k^{i}+\tau ^{{\bf %
(l)}i}{\bf )}$, one can easily obtain the following formula for $\hat{\chi}_{%
{\bf l}}({\bf k},\omega )$ from (\ref{29}):

\begin{eqnarray}
&&\chi _{({\bf l}),\mu }^{\beta }({\bf k},\omega )=\sum\limits_{l}e^{-W_{l}(%
{\bf \tau }^{({\bf l})},{\bf \xi })}e^{-i\tau _{i}^{({\bf l})}\xi
_{l}^{i}}\{- {\displaystyle{\frac{4\pi e^{2}N_{cell} }{m\omega ^{2}}}} f_{l}(%
{\bf \tau }^{({\bf l})})\delta _{\mu }^{\beta }  \nonumber \\
&&+\frac{4\pi N_{cell}}{\omega ^{2}}\frac{2\pi }{2J_{a}(l)+1}
\sum\limits_{\lambda (a)}\rho (\lambda (a))\sum\limits_{L}\left( {%
\displaystyle{\frac{(L+1) }{(2L+1)!!\sqrt{L(L+1)}}}} \right)
^{2}\sum\limits_{J_{b}=\left| L-J_{a}\right| }^{\left| L+J_{a}\right| }\int
\rho (E_{b})dE_{b}  \nonumber \\
&&\times\frac{c^{2}k^{2L}\left| \left\langle a(l);J_{a}(l)\right| \hat{Q}
_{L}(\omega )\left| b(l);J_{b}(l)\right\rangle \right| ^{2}}{E_{a}+\hbar
\omega -E_{b}+D(E_{a}+\hbar \omega )-i\frac{\hbar }{2}\Gamma (E_{a}+\hbar
\omega )}d_{\beta ,\mu }^{L}(\vartheta -\vartheta _{1})\}  \nonumber
\end{eqnarray}

It is worth to note that for ${\bf l}=0$ the condition $\vartheta -\vartheta
_{1}=0$ is true, and as $d_{\beta ,\mu }^{L}(0)=\delta _{\mu }^{\beta }$,
then $\hat{\chi}_{{\bf l=0}}({\bf k},\omega )$ is definitely a scalar value.
For the wave covectors ${\bf k}$ and ${\bf k+\tau }^{{\bf (l)}}$ (with
magnitudes in vicinity) {\bf {\ }}near the Bragg condition the $\vartheta
-\vartheta _{1}=\Delta \theta $ condition takes place, where $\Delta \theta $
\ is the angle width of the Bragg ''table'', i.e. is the value about $
10^{-4}\div 10^{-6}$ radians. Consequently for the wave covectors \-${\bf k}$
and ${\bf k+\tau }^{{\bf (l)}}$ near the Bragg condition we can let $
d_{\beta ,\mu }^{L}(\Delta \theta )=\delta _{\mu }^{\beta }$ with high
precision and can consider $\chi _{({\bf l}),\mu }^{\beta }({\bf k},\omega )$
being the scalar value $\chi _{({\bf l}),\mu }^{\beta }({\bf k},\omega
)=\chi _{({\bf l})}(\omega )\delta _{\mu }^{\beta }$. Then for $\chi _{({\bf %
l})}(\omega )$ we have the final formula

\begin{eqnarray}
&&\chi _{({\bf l})}({\bf \xi },\omega )=\sum\limits_{l=1}^{r}e^{-W_{l}({\bf %
\tau }^{({\bf l})},{\bf \xi })}e^{-i\tau _{i}^{({\bf l})}\xi _{l}^{i}}\{- {%
\displaystyle{\frac{4\pi e^{2}N_{cell} }{m\omega ^{2}}}} f_{l}({\bf \tau }^{(%
{\bf l})})  \nonumber \\
&&+\frac{4\pi N_{cell}}{k}[\int \frac{d\sigma _{l}^{(ion)}}{dE_{b}}\frac{
(E_{a}+\hbar \omega -E_{b}+D(E_{a}+\hbar \omega ))}{(E_{a}+\hbar \omega
-E_{b}+D(E_{a}+\hbar \omega ))^{2}+\frac{\hbar ^{2}\Gamma ^{2}}{4}}
dE_{b}+i\sigma _{l}^{(ion)}(\omega )]\}  \label{30}
\end{eqnarray}
with $\sigma _{l}^{(ion)}(\omega )$ determined by the expression

\begin{eqnarray}
&&\sigma _{l}^{(ion)}(\omega )=\frac{2\pi }{2J_{a}(l)+1}\sum\limits_{\lambda
(a)}\rho (\lambda (a))\sum\limits_{L}\left( {\displaystyle{\frac{(L+1) }{%
(2L+1)!!\sqrt{L(L+1)}}}} \right) ^{2}k^{2L-1} \\
&&\times \sum\limits_{J_{b}=\left| L-J_{a}\right| }^{\left| L+J_{a}\right|
}\int \rho (E_{b})\left| \left\langle a(l);J_{a}(l)\right| \hat{Q}
_{L}(\omega )\left| b(l);J_{b}(l)\right\rangle \right| ^{2}\delta
(E_{a}+\hbar \omega -E_{b}+D(E_{a}+\hbar \omega ))dE_{b}  \nonumber
\end{eqnarray}
and being nothing but the full cross-section of the photo-ionization
(photo-absorption) of the $l$-th atom in the crystal unit cell and

\begin{eqnarray}
&&\frac{d\sigma _{l}^{(ion)}}{dE_{b}}=\frac{2\pi }{2J_{a}(l)+1}
\sum\limits_{\lambda (a)}\rho (\lambda (a))\sum\limits_{L}\left( {%
\displaystyle{\frac{(L+1) }{(2L+1)!!\sqrt{L(L+1)}}}} \right) ^{2}k^{2L-1}
\nonumber \\
&&\times \sum\limits_{J_{b}=\left| L-J_{a}\right| }^{\left| L+J_{a}\right|
}\int \rho (E_{b})\left| \left\langle a(l);J_{a}(l)\right| \hat{Q}
_{L}(\omega )\left| b(l);J_{b}(l)\right\rangle \right| ^{2}  \label{30b}
\end{eqnarray}
is the differential photo-ionization cross-section. These values are
presently rather well analyzed both theoretically and experimentally \cite%
{[15]}. It is worth to underline that while deriving formula (\ref{30}) we
didn't use the condition of the long-wave assumption $ka\ll 1$ with $a$\
being the atom linear size since this condition for the X-ray radiation with
the wavelength about $1\stackrel{\circ }{A}$ is not true. \

\section{ Two-wave dynamic diffraction on the elastically deformed crystal.}

In this part we will discuss the X-ray quanta diffraction on the elastically
deformed crystals. As the X-ray quanta field in crystals can be considered
as a transverse field with high accuracy, the dynamic diffraction on the
deformed crystals in the two-wave approximation according to the formulas (%
\ref{14a}), (\ref{15}) and (\ref{30}) is described by the following
equations:

\begin{eqnarray}
&&{\bf E}({\bf \xi },\omega )={\bf E}_{{\bf 0}}({\bf \xi },\omega
)e^{ik_{i}(\xi ^{i}+u^{i}({\bf \xi }))}+{\bf E}_{{\bf l}}({\bf \xi },\omega
)e^{i[k_{i}+\tau _{i}^{({\bf l})}]\xi ^{i}+ik_{i}u^{i}({\bf \xi })};
\nonumber \\
&&\left[ \frac{1}{k^{2}}\hat{L}({\bf k}({\bf \xi }), {\displaystyle{\frac{%
\partial }{\partial {\bf \xi }}}} )+1\right] {\bf E}_{{\bf 0}}({\bf \xi }%
,\omega )+\sqrt{\left| g({\bf \xi } )\right| }\chi _{({\bf 0})}({\bf \xi }%
,\omega ){\bf E}_{{\bf 0}}({\bf \xi } ,\omega ) +\sqrt{\left| g({\bf \xi }%
)\right| }\chi _{({\bf l})}({\bf \xi } ,\omega ){\bf E}_{{\bf l}}({\bf \xi }%
,\omega )=0  \nonumber \\
&&\left[ \frac{1}{k^{2}}\hat{L}({\bf k}({\bf \xi })+{\bf \tau }^{({\bf l})},
{\displaystyle{\frac{\partial }{\partial {\bf \xi }}}} )+1\right] {\bf E}_{%
{\bf l}}({\bf \xi },\omega )+\sqrt{\left| g({\bf \xi } )\right| }\chi _{(%
{\bf 0})}({\bf \xi },\omega ){\bf E}_{{\bf l}}({\bf \xi } ,\omega )
\nonumber \\
&&+\sqrt{\left| g({\bf \xi })\right| }\chi _{(-{\bf l})}({\bf \xi },\omega )
{\bf E}_{{\bf 0}}({\bf \xi },\omega )=0  \label{31}
\end{eqnarray}

By definition for the elastically deformed crystals the condition $\left| {%
\displaystyle{\frac{\partial u^{i} }{\partial \xi ^{j}}}} \right| \ll 1$
holds true. Then owing to the fact that $\left| \chi _{({\bf 0 },{\bf l}%
)}\right| \thicksim 10^{-4}\div 10^{-6}$, we can assume in (\ref{31} ) $%
\sqrt{\left| g\right| }\chi _{(\pm {\bf l})}\thickapprox \chi _{(\pm {\bf l}%
)}$, $\sqrt{\left| g({\bf \xi })\right| }\chi _{({\bf 0})}\thickapprox \chi
_{({\bf 0})}$. Because the deformation field and accordingly the metric
tensor $g^{lj}$ can be considered to be the constant values in the area with
linear dimensions about the X-ray quanta wavelength, the constant terms
containing $\frac{1}{\sqrt{g}} {\displaystyle{\frac{1 }{k}}} {\displaystyle{%
\frac{\partial (\sqrt{g}g^{lj}) }{\partial \xi ^{l}}}} $ can be neglected.
The terms $\frac{1}{k^{2}}g^{lj} {\displaystyle{\frac{\partial ^{2} }{%
\partial \xi ^{l}\partial \xi ^{j}}}} {\bf E}_{{\bf 0,l}}({\bf \xi },\omega
) $ in the assumption of slowly varying amplitudes can also be neglected. If
one introduces the designations $n_{i}^{({\bf 0})}= {\displaystyle{\frac{%
k_{i} }{k}}} $ for the guiding cosines of wave covector ${\bf k}$ of X-ray
quanta incident on the perfect crystal and $n_{i}^{({\bf l})}= {\displaystyle%
{\frac{k_{i}+\tau _{i} }{\sqrt{({\bf k}+{\bf \tau })^{2}}}}} = {\displaystyle%
{\frac{k_{i}+\tau _{i} }{k\sqrt{1+\alpha }}}} $ for the guiding cosines of
wave covector ${\bf k}$ of X-ray quanta diffracted on the perfect crystal,
then by accounting the assumptions made above and the fact that in the
Lagrange co-ordinate system there takes place the identity $n_{i}^{({\bf 0}%
)}({\bf \xi })g^{ij}n_{j}^{({\bf 0})}({\bf \xi } )=1$whis $n_{i}^{({\bf 0})}(%
{\bf \xi })=n_{\mu }^{({\bf 0})}\left( \delta _{i}^{\mu }+ {\displaystyle{%
\frac{\partial u^{\mu }({\bf \xi }) }{\partial \xi ^{i}}}} \right) $ as the
wave covector ${\bf k}$ guiding cosines in the Lagrange co-ordinate system,
the equations system (\ref{31}) can be sufficiently simplified
\begin{eqnarray}
&&{\bf E}({\bf \xi },\omega )={\bf E}_{{\bf 0}}({\bf \xi },\omega
)e^{ik(n_{i}^{({\bf 0})}\xi ^{i}+n_{i}^{({\bf 0})}u^{i}({\bf \xi }))}+{\bf E}
_{{\bf l}}({\bf \xi },\omega )e^{ik(\sqrt{1+\alpha }n_{i}^{({\bf l})}\xi
^{i}+n_{i}^{({\bf 0})}u^{i}({\bf \xi }))};  \nonumber \\
&&\left[ 2i\frac{1}{k}n_{i}^{({\bf 0})}({\bf \xi })g^{ij} {\displaystyle{%
\frac{\partial }{\partial \xi ^{j}}}} +\chi _{({\bf 0})}({\bf \xi },\omega )%
\right] {\bf E}_{{\bf 0}}({\bf \xi } ,\omega )+\chi _{({\bf l})}({\bf \xi }%
,\omega ){\bf E}_{{\bf l}}({\bf \xi } ,\omega )=0;  \nonumber \\
&&\{2i\frac{1}{k}[\sqrt{1+\alpha }n_{i}^{({\bf l})}+n_{\mu }^{({\bf 0})}%
\frac{\partial u^{\mu }({\bf \xi })}{\partial \xi ^{i}}]g^{ij} {\displaystyle%
{\frac{\partial }{\partial \xi ^{j}}}} +\chi _{({\bf 0})}({\bf \xi },\omega
)+1  \nonumber \\
&&-[\sqrt{1+\alpha }n_{i}^{({\bf l})}+n_{\mu }^{({\bf 0})}\frac{\partial
-u^{\mu }({\bf \xi })}{\partial \xi ^{i}}]g^{lj}[\sqrt{1+\alpha }n_{j}^{(
{\bf l})}+n_{\mu }^{({\bf 0})}\frac{\partial u^{\mu }({\bf \xi })}{\partial
\xi ^{j}}]\}{\bf E}_{{\bf l}}({\bf \xi },\omega )  \nonumber \\
&&+\chi _{(-{\bf l})}({\bf \xi },\omega ){\bf E}_{{\bf 0}}({\bf \xi },\omega
)=0.  \label{32}
\end{eqnarray}
This equations system describes two-wave dynamic diffraction of X-ray quanta
on the deformed crystals and is the system that will be considered as basic
one.

There is Takagi -Taupin equations system that is usually used in the X-ray
optics works. We'll show that the Takagi -Taupin equations system can be
obtained from the system (\ref{32}) after a set of simplifying assumptions.
Let us make the following assumptions:\ 1) $g_{ij}\thickapprox \delta
_{ij}+\delta _{i\mu } {\displaystyle{\frac{\partial u^{\mu }({\bf \xi }) }{%
\partial \xi ^{j}}}} +\delta _{j\mu } {\displaystyle{\frac{\partial u^{\mu }(%
{\bf \xi }) }{\partial \xi ^{i}}}} $, $g^{ij}\thickapprox \delta
^{ij}-\delta ^{i\mu } {\displaystyle{\frac{\partial u^{j}({\bf \xi }) }{%
\partial \xi ^{\mu }}}} -\delta ^{j\mu } {\displaystyle{\frac{\partial u^{i}(%
{\bf \xi }) }{\partial \xi ^{\mu }}}} $ \ (we recall the metric tensor
formula $g_{ij}=\delta _{ij}+\delta _{i\mu } {\displaystyle{\frac{\partial
u^{\mu }({\bf \xi }) }{\partial \xi ^{j}}}} +\delta _{j\mu } {\displaystyle{%
\frac{\partial u^{\mu }({\bf \xi }) }{\partial \xi ^{i}}}} +\delta _{\mu \nu
} {\displaystyle{\frac{\partial u^{\mu }({\bf \xi }) }{\partial \xi ^{i}}}} {%
\displaystyle{\frac{\partial u^{\nu }({\bf \xi }) }{\partial \xi ^{j}}}} $ ,
i.e. we neglect all the values of $o( {\displaystyle{\frac{\partial u^{\mu }(%
{\bf \xi }) }{\partial \xi ^{i}}}} {\displaystyle{\frac{\partial u^{\nu }(%
{\bf \xi }) }{\partial \xi ^{j}}}} )$ order);

2) $n_{i}^{({\bf 0})}({\bf \xi })g^{ij} {\displaystyle{\frac{\partial }{%
\partial \xi ^{j}}}} {\bf E}_{{\bf 0}}({\bf \xi },\omega )\thickapprox
n_{i}^{({\bf 0})}\delta ^{ij} {\displaystyle{\frac{\partial }{\partial \xi
^{j}}}} {\bf E}_{{\bf 0}}({\bf \xi },\omega )$,$[\sqrt{1+\alpha }n_{i}^{(%
{\bf l} )}+n_{\mu }^{({\bf 0})}\frac{\partial u^{\mu }({\bf \xi })}{\partial
\xi ^{i} }]g^{ij} {\displaystyle{\frac{\partial }{\partial \xi ^{j}}}} {\bf E%
}_{{\bf l}}({\bf \xi },\omega )\thickapprox n_{i}^{({\bf l})}\delta ^{ij} {%
\displaystyle{\frac{\partial }{\partial \xi ^{j}}}} {\bf E}_{{\bf l}}({\bf %
\xi },\omega )$;

3) $\chi _{({\bf 0,\pm l})}({\bf \xi },\omega )\thickapprox \chi _{({\bf %
0,\pm l})}(\omega )$, i.e. we neglect the dependence of Debye-Waller factor
on the co-ordinates. Placing these approximated equations into the system of
equations (\ref{32}), we obtain for the slowly varying wave amplitudes

\begin{eqnarray}
&&\left[ 2i\frac{1}{k}n_{i}^{({\bf 0})}\delta ^{ij} {\displaystyle{\frac{%
\partial }{\partial \xi ^{j}}}} +\chi _{({\bf 0})}(\omega )\right] {\bf E}_{%
{\bf 0}}({\bf \xi },\omega )+\chi _{({\bf l})}(\omega ){\bf E}_{{\bf l}}(%
{\bf \xi },\omega )=0;  \nonumber \\
&&\left[ 2i\frac{1}{k}n_{i}^{({\bf l})}\delta ^{ij} {\displaystyle{\frac{%
\partial }{\partial \xi ^{j}}}} +\frac{1}{k}\sqrt{1+\alpha }n_{i}^{({\bf l}%
)}\delta ^{ij}\tau _{\mu } {\displaystyle{\frac{\partial u^{\nu }({\bf \xi }%
) }{\partial \xi ^{j}}}} -\alpha +\chi _{({\bf 0})}(\omega )\right] {\bf E}_{%
{\bf l}}({\bf \xi } ,\omega )  \nonumber \\
&&+\chi _{(-{\bf l})}(\omega ){\bf E}_{{\bf 0}}({\bf \xi },\omega )=0.
\nonumber
\end{eqnarray}
Next, let us introduce vectors of $\pi -$ and $\sigma -$ polarization $
e_{\sigma ,\pi }^{({\bf 0,l})}({\bf \xi }),$\ being standard for the X-ray
optics (it is taken into account that we are applying the Lagrange
co-ordinates system and polarization vectors are the co-ordinate functions).
The values $2i\frac{1}{k}n_{i}^{({\bf 0})}\delta ^{ij}\Gamma _{j\nu }^{\mu
}( {\bf \xi })$ and $2i\frac{1}{k}n_{i}^{({\bf l})}\delta ^{ij}\Gamma _{j\nu
}^{\mu }({\bf \xi })$, where $\Gamma _{j\nu }^{\mu }({\bf \xi })$ is the
Kristoffel symbol, that appear during the polarization vectors
differentiation and can be neglected since the metric tensor remains a
constant value in the area with linear dimensions about the X-ray quanta
wavelength. Then if we assume $\sqrt{1+\alpha }\thickapprox 1$ in the second
equation, these equations look as follows
\begin{eqnarray}
&&\left[ 2i\frac{1}{k}n_{i}^{({\bf 0})}\delta ^{ij} {\displaystyle{\frac{%
\partial }{\partial \xi ^{j}}}} +\chi _{({\bf 0})}(\omega )\right] E_{{\bf 0}%
}({\bf \xi },\omega )+C({\bf \xi })\chi _{({\bf l})}(\omega )E_{{\bf l}}(%
{\bf \xi },\omega )=0;  \nonumber \\
&&\left[ 2i\frac{1}{k}n_{i}^{({\bf l})}\delta ^{ij} {\displaystyle{\frac{%
\partial }{\partial \xi ^{j}}}} +\frac{1}{k}n_{i}^{({\bf l})}\delta
^{ij}\tau _{\mu } {\displaystyle{\frac{\partial u^{\nu }({\bf \xi }) }{%
\partial \xi ^{j}}}} -\alpha +\chi _{({\bf 0})}(\omega )\right] E_{{\bf l}}(%
{\bf \xi },\omega )  \nonumber \\
&&+C({\bf \xi })\chi _{(-{\bf l})}(\omega )E_{{\bf 0}}({\bf \xi },\omega )=0
\label{33}
\end{eqnarray}
where $C({\bf \xi )}$ is the polarization factor: $C_{(\sigma )}({\bf \xi }
)=1$ for $\sigma -$polarization. For $\pi -$ polarization we have a bulky
formula
\[
C_{(\pi )}({\bf \xi })=\frac{n_{i}^{({\bf 0})}({\bf \xi })g^{ij}(\sqrt{
1+\alpha }n_{j}^{({\bf l})}+n_{\mu }^{({\bf 0})} {\displaystyle{\frac{%
\partial u^{\mu }({\bf \xi }) }{\partial \xi ^{j}}}} )}{\sqrt{(\sqrt{%
1+\alpha }n_{i}^{({\bf l})}+n_{\mu }^{({\bf 0})} {\displaystyle{\frac{%
\partial u^{\mu }({\bf \xi }) }{\partial \xi ^{i}}}} )g^{ij}(\sqrt{1+\alpha }%
n_{j}^{({\bf l})}+n_{\nu }^{({\bf 0})} {\displaystyle{\frac{\partial u^{\nu
}({\bf \xi }) }{\partial \xi ^{j}}}} )}}.
\]
As $\left| \chi _{({\bf 0},{\bf l})}\right| \thicksim 10^{-4}\div 10^{-6}$,
we can assume $C_{(\pi )}({\bf \xi })=n_{i}^{({\bf 0})}\delta ^{ij}n_{j}^{(
{\bf l})}$ i.e. we can assume $C_{(\pi )}({\bf \xi })$ to be the same as
that for the perfect crystal. In this case the equations (\ref{33}) for the
slowly changing amplitudes $E_{{\bf 0,l}}({\bf \xi },\omega )$ fully
coincide with Takagi -Taupin equations. It is worth mentioning that in works
mentioned above, in the formula for the X-ray quanta field in the crystal

\[
{\bf E}({\bf \xi },\omega )={\bf E}_{{\bf 0}}({\bf \xi },\omega
)e^{ik(n_{i}^{({\bf 0})}\xi ^{i}+n_{i}^{({\bf 0})}u^{i}({\bf \xi }))}+{\bf E}
_{{\bf l}}({\bf \xi },\omega )e^{ik(\sqrt{1+\alpha }n_{i}^{({\bf l})}\xi
^{i}+n_{i}^{({\bf 0})}u^{i}({\bf \xi }))}
\]
the term $n_{i}^{({\bf 0})}u^{i}({\bf \xi )}$ in the expression for the
passed and diffracted wave phases is left out for some reason and it is the
expression that determines the X-ray quanta focusing by the distorted
crystal both in Laue and in Bragg geometries.

To determine the application field of simplifying assumptions made above and
accordingly to determine the Takagi -Taupin equations application field let
us determine the plane-parallel plate with thickness $h$. Let this plate to
be deformed by the bending momentums $M_{1}$ and $M_{2}$ uniformly
distributed over the plate sides. Then the deformation field can be
introduced by the following way \cite{[16]}:

\begin{eqnarray}
u^{1}({\bf \xi }) &=&-\frac{1}{\rho _{1}}\left[ b_{15}(\xi ^{3}+\frac{h}{2}%
)^{2}+b_{16}\xi ^{2}(\xi ^{3}+\frac{h}{2})+2b_{11}\xi ^{1}(\xi ^{3}+\frac{h}{%
2})\right]   \nonumber \\
&&-\frac{1}{\rho _{2}}\left[ b_{25}(\xi ^{3}+\frac{h}{2})^{2}+b_{26}\xi
^{2}(\xi ^{3}+\frac{h}{2})+2b_{12}\xi ^{1}(\xi ^{3}+\frac{h}{2})\right] ;
\nonumber \\
u^{2}({\bf \xi }) &=&-\frac{1}{\rho _{1}}\left[ b_{14}(\xi ^{3}+\frac{h}{2}%
)^{2}+2b_{12}\xi ^{2}(\xi ^{3}+\frac{h}{2})+2b_{16}\xi ^{1}(\xi ^{3}+\frac{h%
}{2})\right]   \nonumber \\
&&-\frac{1}{\rho _{2}}\left[ b_{24}(\xi ^{3}+\frac{h}{2})^{2}+2b_{22}\xi
^{2}(\xi ^{3}+\frac{h}{2})+b_{26}\xi ^{1}(\xi ^{3}+\frac{h}{2})\right] ;
\nonumber \\
u^{3}({\bf \xi }) &=&\frac{1}{\rho _{1}}\left[ b_{11}(\xi
^{1})^{2}+b_{12}(\xi ^{2})^{2}-b_{13}(\xi ^{3}+\frac{h}{2})^{2}+b_{16}\xi
^{1}\xi ^{2}\right]   \nonumber \\
&&+\frac{1}{\rho _{2}}\left[ b_{12}(\xi ^{1})^{2}+b_{22}(\xi
^{2})^{2}-b_{23}(\xi ^{3}+\frac{h}{2})^{2}+b_{26}\xi ^{1}\xi ^{2}\right] ;
\label{34}
\end{eqnarray}%
where ${\displaystyle{\frac{1}{\rho _{1}}}}={\displaystyle{\frac{M_{1}c_{33}%
}{h^{3}}}};$ ${\displaystyle{\frac{1}{\rho _{2}}}}={\displaystyle{\frac{%
M_{2}c_{33}}{h^{3}}}};$ $b_{\mu \nu }={\displaystyle{\frac{c_{\mu \nu }}{%
c_{33}}}};$ $c_{\mu \nu }$ ($\mu ,\nu =1,\cdots ,6$) is the crystal elastic
constants. The Fig.1 shows this deformed silicon plate for the parameters $%
\rho _{1}=2\times 10^{2}cm$ and $\rho _{2}=5\times 10^{2}cm$ and the
thickness $h=0.1cm$,
\begin{figure}[tbp]
\begin{center}
\epsfig{figure=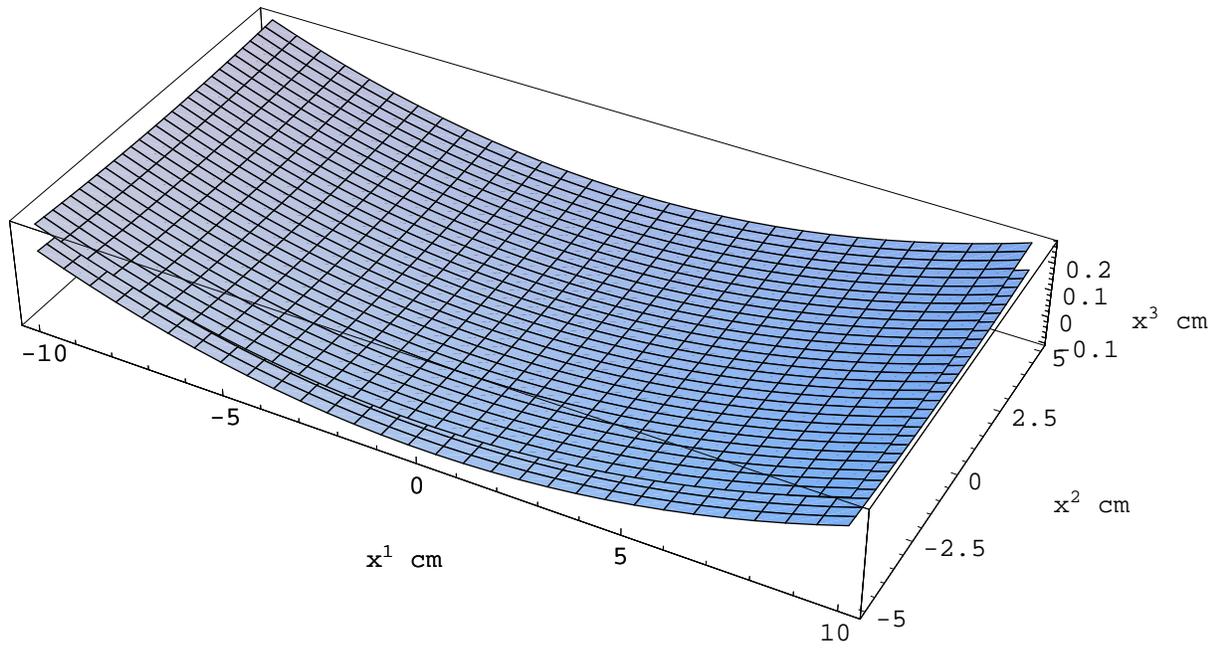}
\end{center}
\caption{ Curved Bragg mirror.}
\label{fig1.eps}
\end{figure}
and Fig.2 shows the{\bf {\ }}average radius of curvature of reflecting
external surface of this plate.
\begin{figure}[tbp]
\begin{center}
\epsfig{figure=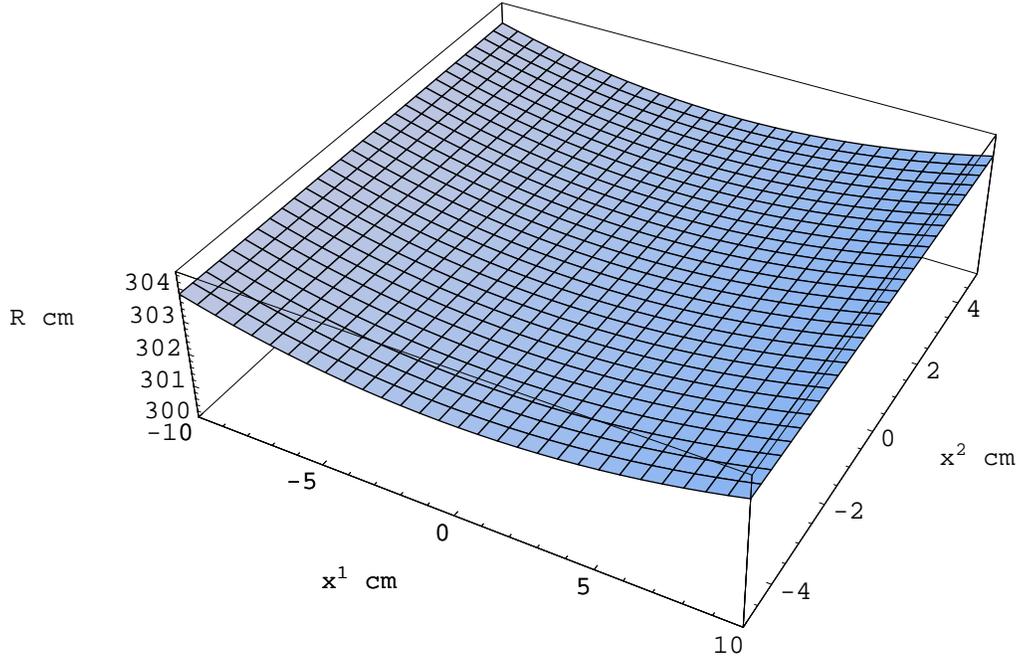}
\end{center}
\caption{The mirror average curvature radius}
\label{fig2.eps}
\end{figure}
As Fig.2 shows, the bent silicon plate average radius of curvature with such
choice of parameters depends on the co-ordinates of a point on the
reflecting surface and is being the value about $3\times 10^{2}cm$. Let us
discuss the most simple situation: the absorption of X-ray quanta passing
through the deformed silicon plate (Bugger law).For this situation {\bf {f}}%
rom equations (\ref{32}{\bf {)}} we obtain (from equations (\ref{32}))

\[
\left[ 2i\frac{1}{k}n_{i}^{({\bf 0})}({\bf \xi })g^{ij}({\bf \xi }) {%
\displaystyle{\frac{\partial }{\partial \xi ^{j}}}} +\chi _{({\bf 0}%
)}(\omega )\right] {\bf E}_{{\bf 0}}({\bf \xi },\omega )=0,
\]
for the simplicity we will neglect the dependence on Debye-Waller factor on
the co-ordinates and from Takagi -Taupin equations (\ref{33})

\[
\left[ 2i\frac{1}{k}n_{i}^{({\bf 0})}\delta ^{ij}{\displaystyle{\frac{%
\partial }{\partial \xi ^{j}}}}+\chi _{({\bf 0})}(\omega )\right] E_{{\bf 0}%
}({\bf \xi },\omega )=0.
\]%
Then solving the first equation by the characteristics method, i.e. by
solving the system of the ordinary differential equations\ ${\displaystyle{%
\frac{d\xi ^{j}(s)}{ds}}}=n_{i}^{({\bf 0})}({\bf \xi })g^{ij}({\bf \xi })$
with the initial conditions $\xi ^{1}(0)=\xi _{0}^{1},\xi ^{2}(0)=\xi
_{0}^{2},$ $\xi ^{3}(0)=0$, where $(\xi _{0}^{1},\xi _{0}^{2},0)$ are the
entrance co-ordinates of the X-ray beam incident on the crystal, we obtain
the following equation for transition function:
\[
T(\xi _{0}^{1},\xi _{0}^{2},h)=e^{-k\mathop{\rm Im}(\chi _{({\bf 0})}(\omega
))s(\xi _{0}^{1},\xi _{0}^{2},h)},
\]%
where $h$ is the perfect crystal length. The X-ray quanta optical path $%
s(\xi _{0}^{1},\xi _{0}^{2},h)$ in the deformed crystal can be obtained by
solving the equation $\xi ^{3}(s,\xi _{0}^{1},\xi _{0}^{2})=h$, where $\xi
^{3}(s,\xi _{0}^{1},\xi _{0}^{2})$ is the solution of the ordinary
differential equations system mentioned above. For the second equation, the
solution i.e. the gating function can be easily obtained and does not differ
from the perfect crystal plane-parallel plate at all.
\[
T=e^{-k\mathop{\rm Im}(\chi _{({\bf 0})}(\omega ))\frac{h}{\sin (\theta )}},
\]%
where $\theta $ is the angle between the X-ray quanta incident on the
crystal and the normal to the perfect crystal entrance plane directed inside
the crystal. Fig.3 show these transition functions for $\theta =\frac{\pi }{%
18},\varphi =\frac{\pi }{4},$ $h=0.01cm$ and quanta energies $15KeV$.
\begin{figure}[tbp]
\begin{center}
\epsfig{figure=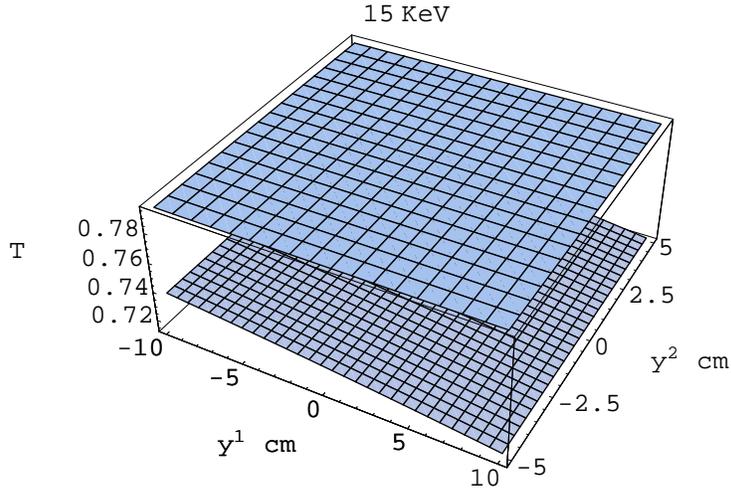}
\end{center}
\caption{ The silicon plate gating function. Here $y^{1}=\protect\xi ^{1}$
and $y^{2}=\protect\xi ^{2}$.}
\label{fig3.eps}
\end{figure}

At the upper surface of this figure is presented the transition function for
the plane-parallel perfect silicon crystal plate and the lower surface
describes the X-ray quanta absorption by the deformed crystal. The
distinction shows itself in the term of the second order and is very
sufficient for such a thin plate. It means that even in such a simple
situation as the absorption of X-ray quanta passing through the deformed
crystal, the Takagi -Taupin equations don't correctly describe the real
experimental situation (this distinction can be neglected only for the
crystal deformation with the average radius of curvature more then $40m$).

To finally define the application range of Takagi -Taupin equations let us
compare two terms:

1) $1-[\sqrt{1+\alpha }n_{i}^{({\bf l})}+n_{\mu }^{({\bf 0})}\frac{\partial
u^{\mu }({\bf \xi })}{\partial \xi ^{i}}]g^{lj}[\sqrt{1+\alpha }n_{j}^{({\bf %
l})}+n_{\mu }^{({\bf 0})}\frac{\partial u^{\mu }({\bf \xi })}{\partial \xi
^{j}}]$\ in the second equation of the equation system (\ref{32}) and

2) $\frac{1}{k}n_{i}^{({\bf l})}\delta ^{ij}\tau _{\mu } {\displaystyle{%
\frac{\partial u^{\nu }({\bf \xi }) }{\partial \xi ^{j}}}} -\alpha $ \ in
the second equation of the Takagi -Taupin equations system ( \ref{33}).
Figure 4 shows the difference

\[
D=1+\alpha -[\sqrt{1+\alpha }n_{i}^{({\bf l})}+n_{\mu }^{({\bf 0})}\frac{%
\partial u^{\mu }({\bf \xi })}{\partial \xi ^{i}}]g^{lj}[\sqrt{1+\alpha }%
n_{j}^{({\bf l})}+n_{\mu }^{({\bf 0})}\frac{\partial u^{\mu }({\bf \xi })}{%
\partial \xi ^{j}}]-\frac{1}{k}n_{i}^{({\bf l})}\delta ^{ij}\tau _{\mu }{%
\displaystyle{\frac{\partial u^{\nu }({\bf \xi })}{\partial \xi ^{j}}}}
\]%
between these values for the deformed silicon plate shown on the
Fig.4-6 for the energies $5KeV,10KeV$ and $15KeV$ of the quanta.

\begin{figure}[tbp]
\begin{center}
\epsfig{figure=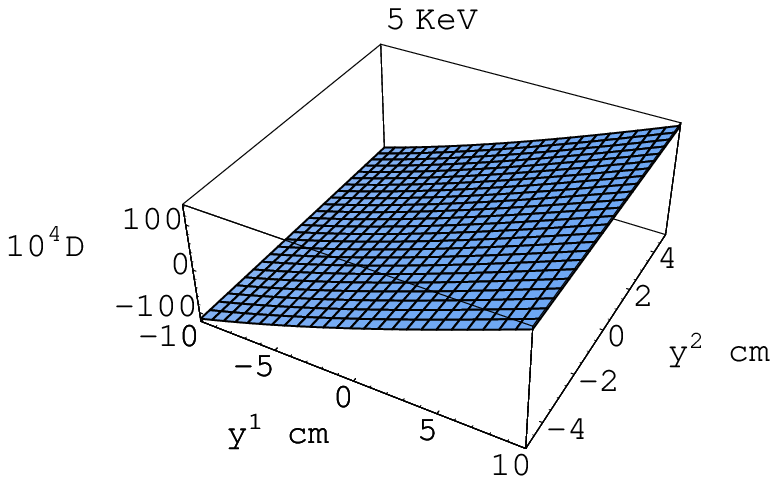}
\end{center}
\caption{ The value $D$ for the quanta with energy $5KeV$ . Here $y^{1}=%
\protect\xi ^{1}$ and $y^{2}=\protect\xi ^{2}$.}
\label{fig4a.eps}
\end{figure}

\begin{figure}[tbp]
\begin{center}
\epsfig{figure=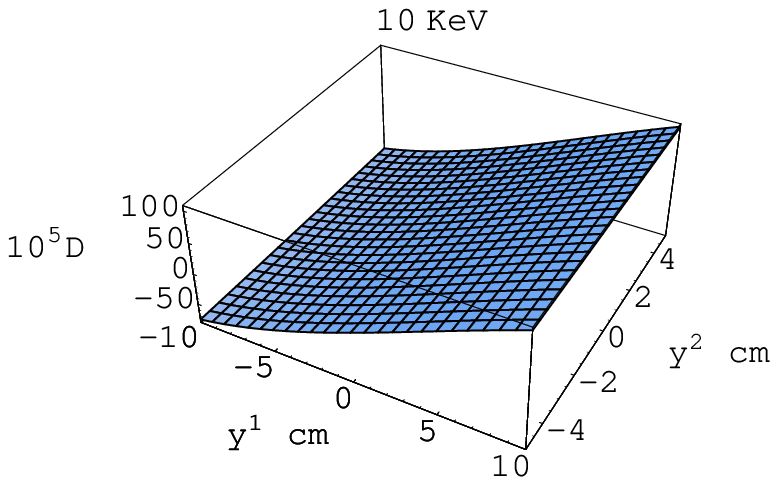}
\end{center}
\caption{ The value $D$ for the quanta with energy $10KeV$. Here $y^{1}=%
\protect\xi ^{1}$ and $y^{2}=\protect\xi ^{2}$.}
\label{fig4b.eps}
\end{figure}

\begin{figure}[tbp]
\begin{center}
\epsfig{figure=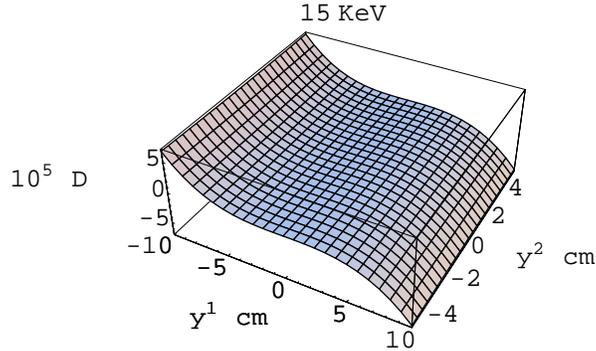}
\end{center}
\caption{ The value $D$ for the quanta with energy $15KeV$. Here $y^{1}=%
\protect\xi ^{1}$ and $y^{2}=\protect\xi ^{2}$.}
\label{fig4c.eps}
\end{figure}

\bigskip

If one notify that for energies $5KeV,10KeV$ and $15KeV$ we have $%
\mathop{\rm Re}(\chi _{({\bf 0})}(\omega ))=-0.394922\times 10^{-4}$,$%
-0.976232\times 10^{-5}$ and $\ -0.431670\times 10^{-5}$ accordingly, this
figure shows that by accepting the simplifying assumptions in the equations (%
\ref{32}) leading to the Takagi -Taupin equations we reject the values that
are 10 times greater than the values we keep (for the crystals with the
average radius of curvature less than $10m$) while obtaining these
equations. This difference shown in Fig.4 becomes less then $\mathop{\rm Re}%
(\chi _{({\bf 0})}(\omega ))$ for the deformations with the curvature
average radius more then $40m$, i.e. very slightly deformed crystals. That
is why one should use exactly the equations system (\ref{32}) for the
theoretical analysis. This system is a little bit more complicated but is
suitable for the description of the X-ray optics devices being met with in
practice.

\section{ \ \ \ \ \ \ \ \ \ \ Bend Bragg mirror.}

The equations (\ref{32}) ( equations for the slowly varying amplitudes of
the field of X-ray quanta in the two-wave dynamic diffraction on the
deformed crystals) for the arbitrary deformation field can be solved in
general only numerically. In the article we underlined more than once that
the X-ray optics can be considered as the geometrical optics\ from the point
of view of general electrodynamics. That is why to obtain the coefficient of
Bragg reflection from the distorted Bragg mirror we will use the geometrical
optics locality principle (see for ex. \cite{[17]}) that evidently was first
time introduced in electrodynamics by V.A. Fock \cite{[18]} who investigated
the reflection of the electromagnetic wave with smooth arbitrary shape front
on the smooth arbitrary shaped surface. According to the locality principle
the wave reflection in the every mirror point acts like if the incident wave
is being smooth one and the curvilinear surface in this point is replaced by
the tangent plane. But it is necessary to take into account that the
incident and the reflection angle change their values from point to point on
this surface. Consequently the locality principle allows us to use the
reflection coefficients obtained for the plane waves and plane surfaces, and
in particular the Bragg reflection coefficient for the diffraction X-ray
optics as the primary approximation. The Bragg reflection coefficient in the
case of the plane wave reflecting from the parallel-sided perfect crystal
thick plate is determined by formula:

\begin{equation}
{\cal R}=\frac{C\chi _{({\bf -l})}(\omega )}{\chi _{({\bf 0})}(\omega
)(1+\beta )- {\displaystyle{\frac{\alpha }{2}}} \pm \sqrt{\left( {%
\displaystyle{\frac{\chi _{({\bf 0})}(\omega ) }{2}}} (1+\beta )- {%
\displaystyle{\frac{\alpha }{2}}} \right) ^{2}-C^{2}\beta \chi ^{2}(\omega )}%
}  \label{35}
\end{equation}
where $C$ is the polarization factor mentioned above, $\beta =- {%
\displaystyle{\frac{n_{3}^{({\bf l})} }{n_{3}^{({\bf 0})}}}} $ is the
asymmetry parameter, $\chi ^{2}(\omega )=\chi _{({\bf -l})}(\omega )\chi _{(%
{\bf l})}(\omega )$, $\alpha = {\displaystyle{\frac{({\bf k}+{\bf \tau }%
)^{2}-k^{2} }{k^{2}}}} $ is the parameter of precise Bragg condition
deviation.

Let the X-ray quanta wave packet incident on the Bragg mirror with arbitrary
smooth deformation field ${\bf u}({\bf \xi })$ has the electric vector for
quanta with frequency $\omega $ ( or the energy $\varepsilon $ ) determined
in the Lagrange co-ordinates system by the equation

\[
{\bf E}({\bf \xi },\omega )={\bf E}^{(0)}({\bf \xi },\omega
)e^{ik(n_{i}^{(0)}\xi ^{i}+n_{i}^{(0)}u^{i}({\bf \xi })+\Psi _{0}({\bf \xi }
,\varepsilon ))},
\]
where $n_{i}^{(0)}$\ are wave vector guiding cosines and ${\bf E}^{(0)}({\bf %
\xi },\omega )$ and $\Psi _{0}({\bf \xi },\varepsilon )$ are the slowly
changing amplitude and phase of the X-ray quanta incident on the mirror.
Then according to the locality principle of geometrical optics this
reflection coefficient of the wave packet for distorted Bragg mirror has the
functional form like that for the plane wave and plane mirror. But it is
necessary to take into account that the values $\chi _{({\bf 0,\pm l}
)},C,\beta $ \ and $\alpha $ for the distorted Bragg mirror change from
point to point on this mirror surface. If the dependence $\chi _{({\bf 0,\pm
l})}$ on co-ordinates is determined by formula (\ref{30}), the dependencies
of $C$,$\beta $ \ and $\alpha $ on the co-ordinates can be defined by the
following ideas. The wave covector components of the X-quanta incident on
the distorted Bragg mirror in the Lagrange co-ordinate system to be
determined by the equation

\[
k_{i}(\xi ^{1},\xi ^{2})=k\left[ n_{i}^{({\bf 0})}+n_{\mu }^{({\bf 0}
)}\left( \frac{\partial u^{\mu }({\bf \xi })}{\partial \xi ^{i}}\right)
_{\xi ^{3}=0}+\left( \frac{\partial \Psi _{0}({\bf \xi },\varepsilon )}{
\partial \xi ^{i}}\right) _{\xi ^{3}=0}\right] .
\]
Then the wave covector X-ray quanta components in the Lagrange co-ordinates
system having diffracted on this mirror are determined by formula

\[
k_{i}(\xi ^{1},\xi ^{2})+\tau _{i}=k\left[ \sqrt{1+\alpha }n_{i}^{({\bf l}
)}+n_{\mu }^{({\bf 0})}\left( \frac{\partial u^{\mu }({\bf \xi })}{\partial
\xi ^{i}}\right) _{\xi ^{3}=0}+\left( \frac{\partial \Psi _{0}({\bf \xi }
,\varepsilon )}{\partial \xi ^{i}}\right) _{\xi ^{3}=0}\right] .
\]
According to the definition we have for parameters $\alpha (\xi ^{1},\xi
^{2}),\beta (\xi ^{1},\xi ^{2})$ and $C(\xi ^{1},\xi ^{2})$

\begin{eqnarray}
&&\alpha (\xi ^{1},\xi ^{2})=  \nonumber \\
&&\frac{2k\tau _{i}g^{ij}(\xi ^{1},\xi ^{2})n_{j}^{({\bf 0})}(\xi ^{1},\xi
^{2})+2k\tau _{i}g^{ij}(\xi ^{1},\xi ^{2})\left( \frac{\partial \Psi _{0}(
{\bf \xi },\varepsilon )}{\partial \xi ^{j}}\right) _{\xi ^{3}=0}+\tau
_{i}g^{ij}(\xi ^{1},\xi ^{2})\tau _{j}}{k^{2}\left[ 1+2n_{i}^{({\bf 0})}(\xi
^{1},\xi ^{2})g^{ij}(\xi ^{1},\xi ^{2})\left( \frac{\partial \Psi _{0}({\bf %
\xi },\varepsilon )}{\partial \xi ^{j}}\right) _{\xi ^{3}=0}+\left( \frac{
\partial \Psi _{0}({\bf \xi },\varepsilon )}{\partial \xi ^{i}}\right) _{\xi
^{3}=0}g^{ij}(\xi ^{1},\xi ^{2})\left( \frac{\partial \Psi _{0}({\bf \xi }
,\varepsilon )}{\partial \xi ^{j}}\right) _{\xi ^{3}=0}\right] };  \nonumber
\\
&&\beta (\xi ^{1},\xi ^{2})=-\frac{n_{3}^{({\bf l})}(\xi ^{1},\xi ^{2})}{
n_{3}^{({\bf 0})}(\xi ^{1},\xi ^{2})};  \nonumber \\
&&C(\xi ^{1},\xi ^{2})=n_{i}^{({\bf 0})}(\xi ^{1},\xi ^{2})g^{ij}(\xi
^{1},\xi ^{2})n_{j}^{({\bf l})}(\xi ^{1},\xi ^{2}),  \label{36}
\end{eqnarray}
where $n_{i}^{({\bf 0})}(\xi ^{1},\xi ^{2})$ and $n_{j}^{({\bf l})}(\xi
^{1},\xi ^{2})$ are the wave covectors\ guiding cosines of X-ray quanta
falling and diffracted on the distorted Bragg mirror surface in the Lagrange
co-ordinates system:

\ \ \ \ \ \ \ \ \ \ \ \ \ \ \ \ \ \ \ \ \ \
\begin{eqnarray}
&&n_{i}^{({\bf 0})}(\xi ^{1},\xi ^{2})=n_{\mu }^{({\bf 0})}\left[ \delta
_{i}^{\mu }+\left( \frac{\partial u^{\mu }({\bf \xi })}{\partial \xi ^{i}}
\right) _{\xi ^{3}=0}\right] ;  \nonumber \\
&&n_{i}^{({\bf l})}(\xi ^{1},\xi ^{2})=\frac{\left[ \sqrt{1+\alpha }n_{i}^{(
{\bf l})}+n_{\mu }^{({\bf 0})}\left( \frac{\partial u^{\mu }({\bf \xi })}{
\partial \xi ^{i}}\right) _{\xi ^{3}=0}+\left( \frac{\partial \Psi _{0}({\bf %
\xi },\varepsilon )}{\partial \xi ^{i}}\right) _{\xi ^{3}=0}\right] }{l(\xi
^{1},\xi ^{2})},  \nonumber \\
&&l(\xi ^{1},\xi ^{2})=\sqrt{\left[ \sqrt{1+\alpha }n_{i}^{({\bf l})}+n_{\mu
}^{({\bf 0})}\left( \frac{\partial u^{\mu }({\bf \xi })}{\partial \xi ^{i}}
\right) _{\xi ^{3}=0}+\left( \frac{\partial \Psi _{0}({\bf \xi },\varepsilon
)}{\partial \xi ^{i}}\right) _{\xi ^{3}=0}\right] g^{ij}(\xi ^{1},\xi ^{2})}
\nonumber \\
&&\times\sqrt{\left[ \sqrt{1+\alpha }n_{j}^{({\bf l})}+n_{\mu }^{({\bf 0}
)}\left( \frac{\partial u^{\mu }({\bf \xi })}{\partial \xi ^{j}}\right)
_{\xi ^{3}=0}+\left( \frac{\partial \Psi _{0}({\bf \xi },\varepsilon )}{
\partial \xi ^{j}}\right) _{\xi ^{3}=0}\right] }  \nonumber
\end{eqnarray}

Consequently the Bragg reflection coefficient of the distorted mirror in
this approximation looks like

\begin{eqnarray}
{\cal R}(\xi ^{1},\xi ^{2}) &=&\frac{C\chi _{({\bf -l})}(\xi ^{1},\xi
^{2};\omega )}{\chi _{({\bf 0})}(\xi ^{1},\xi ^{2};\omega )(1+\beta )- {%
\displaystyle{\frac{\alpha (\xi ^{1},\xi ^{2}) }{2}}} \pm \Delta (\xi
^{1},\xi ^{2})}  \nonumber \\
\Delta (\xi ^{1},\xi ^{2}) &=&\sqrt{\left( {\displaystyle{\frac{\chi _{({\bf %
0})}(\xi ^{1},\xi ^{2};\omega ) }{2}}} (1+\beta )- {\displaystyle{\frac{%
\alpha (\xi ^{1},\xi ^{2}) }{2}}} \right) ^{2}-C^{2}\beta \chi ^{2}(\xi
^{1},\xi ^{2};\omega )}  \label{37}
\end{eqnarray}
Here $\chi _{({\bf 0,\pm l})}(\xi ^{1},\xi ^{2};\omega )=\left( \chi _{({\bf %
0,\pm l})}({\bf \xi };\omega )\right) _{\xi ^{3}=0},\chi ^{2}(\xi ^{1},\xi
^{2};\omega )=\chi _{({\bf l})}(\xi ^{1},\xi ^{2};\omega )\chi _{({\bf -l}%
)}(\xi ^{1},\xi ^{2};\omega )$. The asymmetry parameter $\beta $ and the
polarization factor defined by formulas (\ref{36}) can be considered the
same as for the plane mirror in the first approximation. If besides we
neglect the Debye-Waller dependency on the co-ordinates, the formula (\ref%
{37}) for the Bragg coefficient is extremely simplified:

\
\begin{eqnarray}
&&{\cal R}(\xi ^{1},\xi ^{2})=  \nonumber \\
&=&\frac{C\chi _{({\bf -l})}(\omega )}{\chi _{({\bf 0})}(\omega )(1+\beta )-
{\displaystyle{\frac{\alpha (\xi ^{1},\xi ^{2}) }{2}}} \pm \sqrt{\left( {%
\displaystyle{\frac{\chi _{({\bf 0})}(\omega ) }{2}}} (1+\beta )- {%
\displaystyle{\frac{\alpha (\xi ^{1},\xi ^{2}) }{2}}} \right)
^{2}-C^{2}\beta \chi ^{2}(\omega )}}.  \label{38}
\end{eqnarray}
The fact, that the distorted Bragg mirror with rather big average radius of
curvature can be represented as a set of plane Bragg mirrors turned
relatively to each other, is widely used in X-ray optics systems designing.
Nevertheless we haven't found the formulas (\ref{37}-\ref{38}) introducing
this representation (or their analogues) in other papers. The linear
dimensions of the area participating in the forming of the diffracted field
in the point $(\xi ^{1},\xi ^{2})$ on the crystal surface are about the
X-ray quanta crystal absorption length $l_{abs}\thicksim 10^{-2}\div
10^{-3}cm$. If the deformation field in this area can be considered as the
constant value, the formulas (\ref{37}-\ref{38}) are quite applicable.

Once the Bragg reflection coefficient is obtained, we can define the field
of X-ray quanta diffracted on the distorted Bragg mirror reflecting surface

\begin{eqnarray}
E_{\pi ,\sigma }(\xi ^{1},\xi ^{2};\omega ) &=&{\cal R}(\xi ^{1},\xi
^{2})E_{\pi ,\sigma }^{(0)}(\xi ^{1},\xi ^{2};\omega )e^{ik\Phi _{({\bf l}
)}(\xi ^{1},\xi ^{2})};  \nonumber \\
\Phi _{({\bf l})}(\xi ^{1},\xi ^{2}) &=&\left[ \sqrt{1+\alpha }n_{i}^{({\bf %
l })}+n_{i}^{(0)}u^{i}({\bf \xi })+\Psi _{0}({\bf \xi },\varepsilon )\right]
\label{39}
\end{eqnarray}

The equations (\ref{37}-\ref{38}) and (\ref{39}) solve the problem of X-ray
quanta two-wave dynamic diffraction on the elastically deformed crystal in
the Bragg geometry, i.e. on the incurved Bragg mirror. The only problem
remaining is the X-ray quanta propagation and focusing in vacuum (or in the
air), i.e. the Hemholts equation solution.

\[
(\Delta +k^{2})E({\bf r},\omega )=0
\]
with the assumption that on the distorted Bragg mirror surface there is the
X-ray quanta field $A^{(0)}({\bf r},\omega )e^{ikS_{0}({\bf r})}$ defined
with the amplitude \ $A^{(0)}({\bf r},\omega )={\cal R}(\xi ^{1},\xi
^{2})E^{(0)}(\xi ^{1},\xi ^{2};\omega )$ and phase $S_{0}({\bf r}
,\varepsilon )=\Phi _{({\bf l})}(\xi ^{1},\xi ^{2},\varepsilon )$. Beyond
the caustics this task can be solved by geometrical optics methods \cite%
{[17]} leading to the formula

\begin{equation}
E({\bf r},\omega )={\cal R}(\xi ^{1},\xi ^{2})E^{0}(\xi ^{1},\xi ^{2};\omega
)\sqrt{\frac{R_{1}R_{2}}{(S-R_{1})(S-R_{2})}}e^{ikS_{0}({\bf r})+ikS({\bf r}
,\xi ^{1},\xi ^{2})}  \label{40}
\end{equation}
where
\begin{equation}
S({\bf r},\xi ^{1},\xi ^{2})=\sqrt{(x^{1}-\xi ^{1}-u^{1}(\xi ^{1},\xi
^{2},0))^{2}+(x^{2}-\xi ^{2}-u^{2}(\xi ^{1},\xi
^{2},0))^{2}+(x^{3}-u^{3}(\xi ^{1},\xi ^{2},0))^{2}}  \label{40a}
\end{equation}
is the distance from the Bragg mirror surface point with parameters $(\xi
^{1},\xi ^{2})$ to the point with co-ordinates $(x^{1},x^{2},x^{3})$, and $
R_{1}$ and $R_{2}$ are the curvature radiuses of the wave diffracted on the
Bragg mirror surface defined by the equations

\ \
\begin{eqnarray}
R_{1,2} &=&\frac{-b\pm \sqrt{b^{2}-4ac}}{2a};  \nonumber \\
a &=&\delta ^{ijl}N_{i}(\xi ^{1},\xi ^{2})\frac{\partial N_{j}(\xi ^{1},\xi
^{2})}{\partial \xi ^{1}}\frac{\partial N_{l}(\xi ^{1},\xi ^{2})}{\partial
\xi ^{2}};  \nonumber \\
b &=&\delta ^{ijl}N_{i}(\xi ^{1},\xi ^{2})\frac{\partial N_{j}(\xi ^{1},\xi
^{2})}{\partial \xi ^{1}}\left( \frac{\partial x_{l}({\bf \xi })}{\partial
\xi ^{2}}\right) _{\xi ^{3}=0}-\delta ^{ijl}N_{i}(\xi ^{1},\xi ^{2})\frac{
\partial N_{j}(\xi ^{1},\xi ^{2})}{\partial \xi ^{2}}\left( \frac{\partial
x_{l}({\bf \xi })}{\partial \xi ^{1}}\right) _{\xi ^{3}=0};  \nonumber \\
c &=&\delta ^{ijl}N_{i}(\xi ^{1},\xi ^{2})\left( \frac{\partial x_{j}({\bf %
\xi })}{\partial \xi ^{1}}\right) _{\xi ^{3}=0}\left( \frac{\partial x_{l}(
{\bf \xi })}{\partial \xi ^{2}}\right) _{\xi ^{3}=0}.  \label{41}
\end{eqnarray}

Here $\delta ^{ijl}$ is the fully asymmetrical tensor of the third rank ($
\delta ^{123}=1$), $N_{i}(\xi ^{1},\xi ^{2})$ is the normal vector to the
X-ray quanta wave front on the Bragg mirror surface in the Euler
co-ordinates system defined by the formula

\begin{eqnarray}
&&N_{j}(\xi ^{1},\xi ^{2})=\frac{\left( \hat{J}^{-1}(\xi ^{1},\xi
^{2})\right) _{j}^{i}\left[ \sqrt{1+\alpha }n_{i}^{({\bf l})}+n_{\mu }^{(
{\bf 0})}\left( \frac{\partial u^{\mu }({\bf \xi })}{\partial \xi ^{i}}
\right) _{\xi ^{3}=0}+\left( \frac{\partial \Psi _{0}({\bf \xi },\varepsilon
)}{\partial \xi ^{i}}\right) _{\xi ^{3}=0}\right] }{L(\xi ^{1},\xi ^{2})};
\nonumber \\
&&\left( L(\xi ^{1},\xi ^{2})\right) ^{2}=\left( \hat{J}^{-1}(\xi ^{1},\xi
^{2})\right) _{j}^{i}\left[ \sqrt{1+\alpha }n_{i}^{({\bf l})}+n_{\mu }^{(
{\bf 0})}\left( \frac{\partial u^{\mu }({\bf \xi })}{\partial \xi ^{i}}
\right) _{\xi ^{3}=0}+\left( \frac{\partial \Psi _{0}({\bf \xi },\varepsilon
)}{\partial \xi ^{i}}\right) _{\xi ^{3}=0}\right]  \nonumber \\
&&\times \delta ^{jh}\left( \hat{J}^{-1}(\xi ^{1},\xi ^{2})\right) _{h}^{l} %
\left[ \sqrt{1+\alpha }n_{l}^{({\bf l})}+n_{\mu }^{({\bf 0})}\left( \frac{
\partial u^{\mu }({\bf \xi })}{\partial \xi ^{l}}\right) _{\xi
^{3}=0}+\left( \frac{\partial \Psi _{0}({\bf \xi },\varepsilon )}{\partial
\xi ^{l}}\right) _{\xi ^{3}=0}\right]  \label{42}
\end{eqnarray}
where $\hat{J}^{-1}(\xi ^{1},\xi ^{2})$ is the matrix inverse to the Jacob
matrix $\hat{J}(\xi ^{1},\xi ^{2})=\left\| \delta _{j}^{i}+ {\displaystyle{%
\frac{\partial u^{i}({\bf \xi }) }{\partial \xi ^{j}}}} \right\| _{\xi
^{3}=0}$ \ and $x_{l}({\bf \xi )=\xi }^{l}+u^{l}({\bf \xi })$ are the Euler
co-ordinates of the deformed crystal points.

In the points where the equations $S=R_{1,2}$ are fulfilled, the
geometric-optical solution (40) turns into the infinity (has a singularity).
These points are called the focal ones and the set of the focal points is
called the caustic.\ Consequently the multitude called the caustic is
determined by the equations

\begin{equation}
x_{(f)1,2}^{j}=\left( \xi ^{j}+u^{j}({\bf \xi })\right) _{\xi
^{3}=0}+R_{1,2}(\xi ^{1},\xi ^{2})\delta ^{ji}N_{i}(\xi ^{1},\xi ^{2})
\label{43}
\end{equation}

As the example illustrating formulas (\ref{37})-(\ref{43}) we will consider
a simple situation when the plane waves packet of X-ray quanta with parallel
vectors and energies in the interval from $11KeV$ to $17KeV$ incident on the
plane-parallel silicon plate with thickness $h=0.1cm$ and reflection vector $%
{\bf \tau }(1,1,1)$ perpendicular to the plate entrance surface.
We will consider the crystal plate oriented in such the way that
the Bragg condition precisely holds for the X-ray quanta with
energies $15KeV$. Next this plate is distorted by momentums
$M_{1}$ and $M_{2}$ uniformly distributed over the plate sides in
such a way that the deformations field is determined by the
formulas (\ref{34}) from the previous part and the distorted Bragg
mirror is the result of this deformation shown on Fig.1. Since by
accepted condition the packet of plane waves is incident on this
mirror, one would let $\Psi _{0}({\bf \xi },\varepsilon )=0$ in
formulas (\ref{37})-(\ref{43}) in this situation. The angle
between the diffraction plate and the co-ordinate plate $(\xi
^{1},\xi ^{3})$ will be denoted as $\varphi $. Fig.7-10 shows the
graphs obtained with the help of formula (\ref{38}) for the module
of coefficient
of Bragg reflection from this distorted mirror for X-ray quanta energies $%
11KeV,$ $13KeV,$ $15KeV$ and $17KeV$ with $\varphi =0$.
\begin{figure}[tbp]
\begin{center}
\epsfig{figure=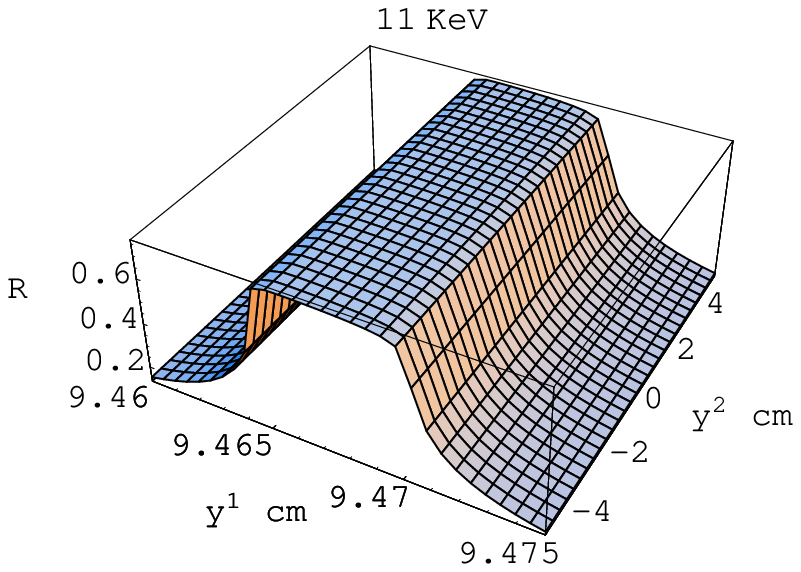}
\end{center}
\caption{ Bragg reflection coefficient. Here $y^{1}=\protect\xi ^{1}$ and $%
y^{2}=\protect\xi ^{2}$. }
\label{fig5a.eps}
\end{figure}

\begin{figure}[tbp]
\begin{center}
\epsfig{figure=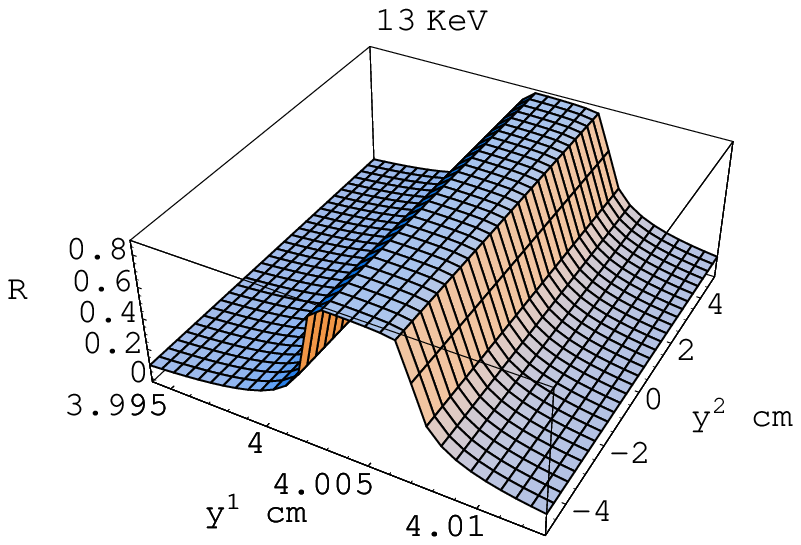}
\end{center}
\caption{ Bragg reflection coefficient. Here $y^{1}=\protect\xi ^{1}$ and $%
y^{2}=\protect\xi ^{2}$.}
\label{fig5b.eps}
\end{figure}

\begin{figure}[tbp]
\begin{center}
\epsfig{figure=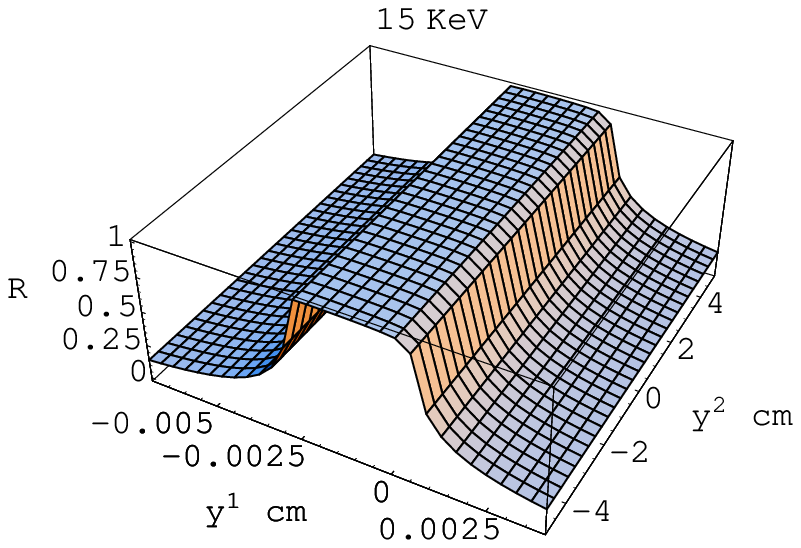}
\end{center}
\caption{ Bragg reflection coefficient. Here $y^{1}=\protect\xi ^{1}$ and $%
y^{2}=\protect\xi ^{2}$.}
\label{fig5c.eps}
\end{figure}

\begin{figure}[tbp]
\begin{center}
\epsfig{figure=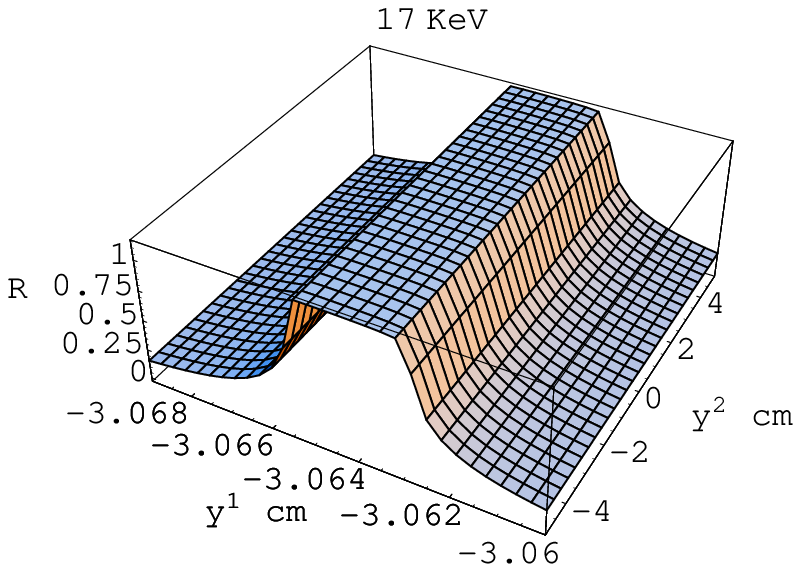}
\end{center}
\caption{ Bragg reflection coefficient. Here $y^{1}=\protect\xi ^{1}$ and $%
y^{2}=\protect\xi ^{2}$.}
\label{fig5d.eps}
\end{figure}

We remind that the caustics (the focal point multitude) of X-ray
quanta diffracted on the Bragg mirror is defined exclusively by
the X-ray radiation field and does not depend on the reflected
quanta intensity distribution (Bragg radiation coefficient) on the
crystal surface. By Fig.11-14 we give the caustics for X-ray
quanta energies $11KeV$, $13KeV$, $15KeV$ and $17KeV$ obtained
from formulas (\ref{42}-\ref{43}) for the whole Bragg mirror
(without accounting the intensity distribution of quanta reflected
on the whole surface of the mirror, i.e. without accounting the
Bragg reflection coefficient dependency on the co-ordinates of the
point on the mirror surface).
\begin{figure}[tbp]
\begin{center}
\epsfig{figure=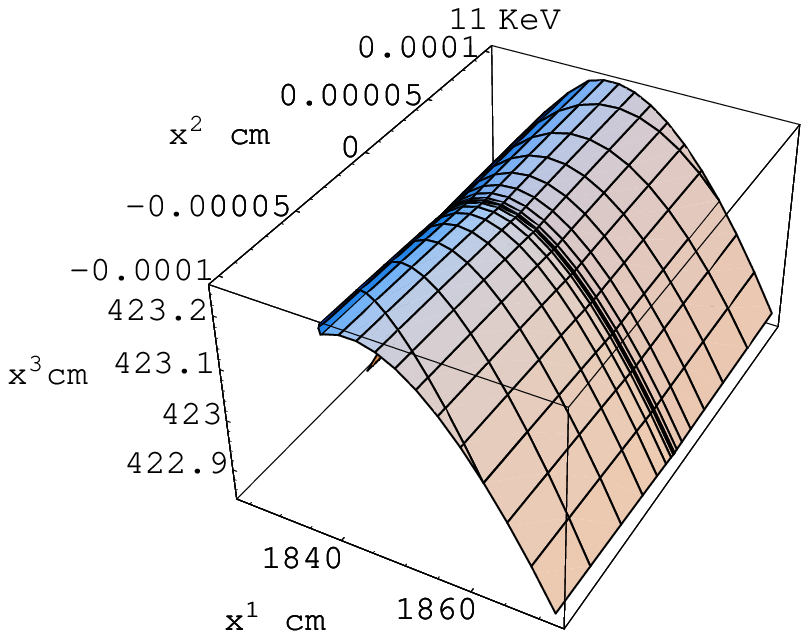}
\end{center}
\caption{ Diffracted quanta caustic.}
\label{fig6a.eps}
\end{figure}

\begin{figure}[tbp]
\begin{center}
\epsfig{figure=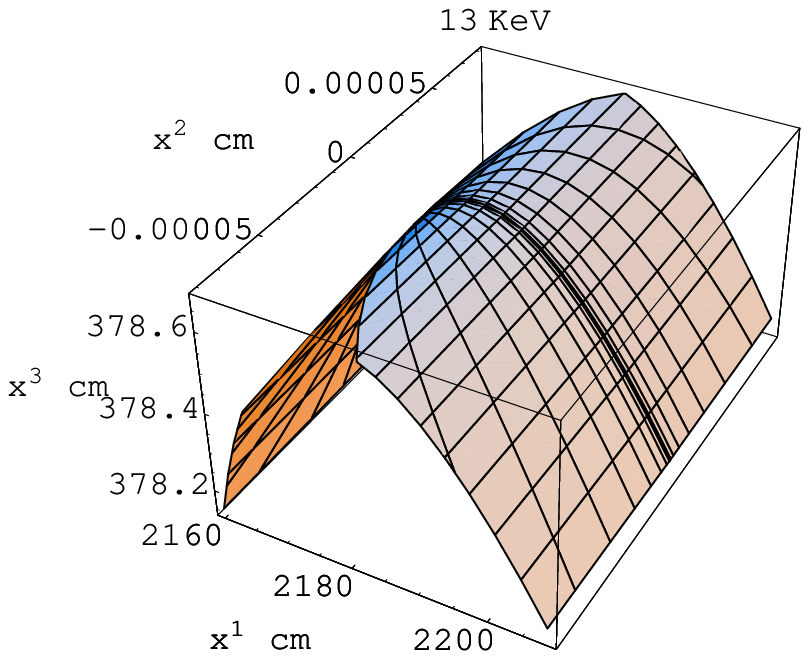}
\end{center}
\caption{ Diffracted quanta caustic.}
\label{fig6b.eps}
\end{figure}

\begin{figure}[tbp]
\begin{center}
\epsfig{figure=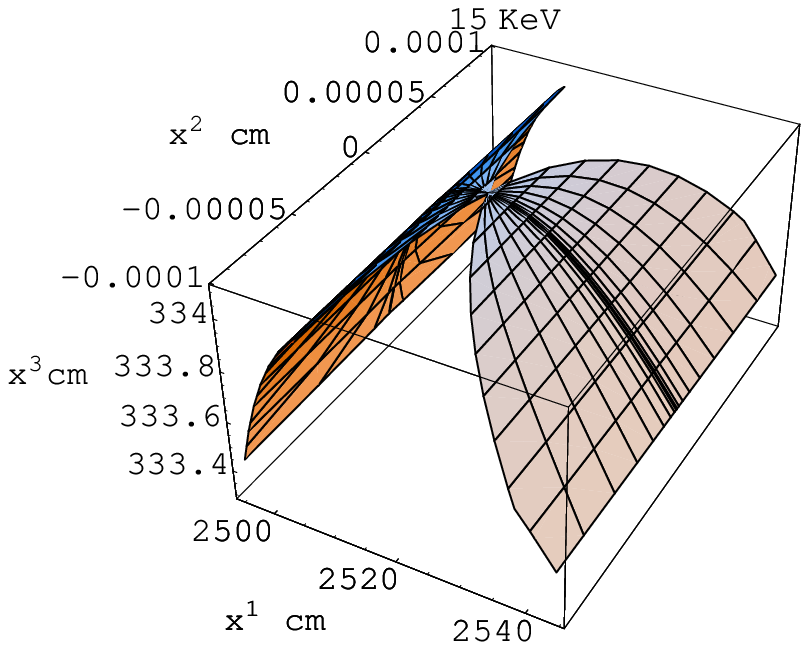}
\end{center}
\caption{ Diffracted quanta caustic.}
\label{fig6c.eps}
\end{figure}

\begin{figure}[tbp]
\begin{center}
\epsfig{figure=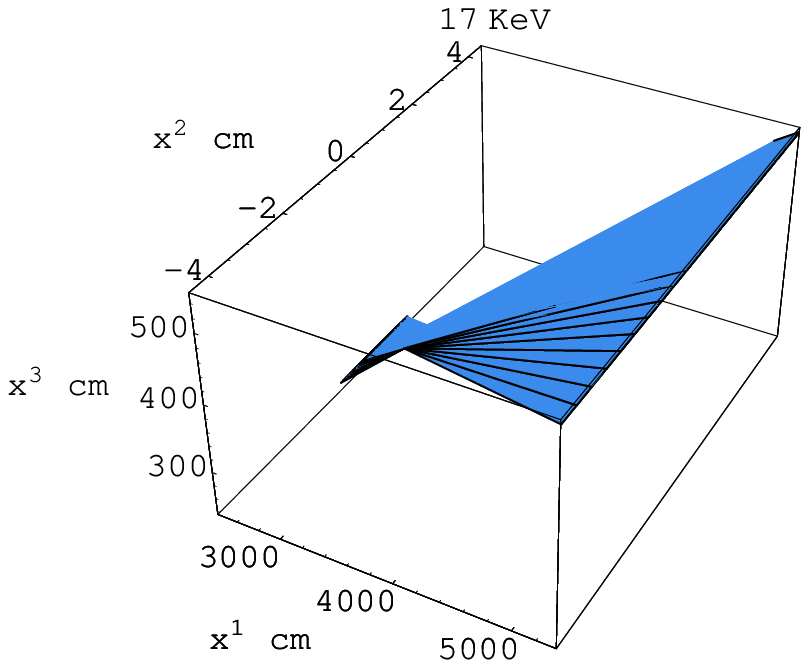}
\end{center}
\caption{ Diffracted quanta caustic.}
\label{fig6d.eps}
\end{figure}

It is clear that in experiment there will be presented only the
part of this caustic surface which X-ray quanta are coming at,
once reflected from the mirror area with the Bragg reflection
coefficient sufficiently differing from null. These areas for our
example are shown on fig.5. Fig.15-18 shows
caustics (focal points manifold) obtained also with the help of formulas ( %
\ref{41}-\ref{43}) but taking into account the dependency of Bragg
reflection coefficients on the co-ordinates.
\begin{figure}[tbp]
\begin{center}
\epsfig{figure=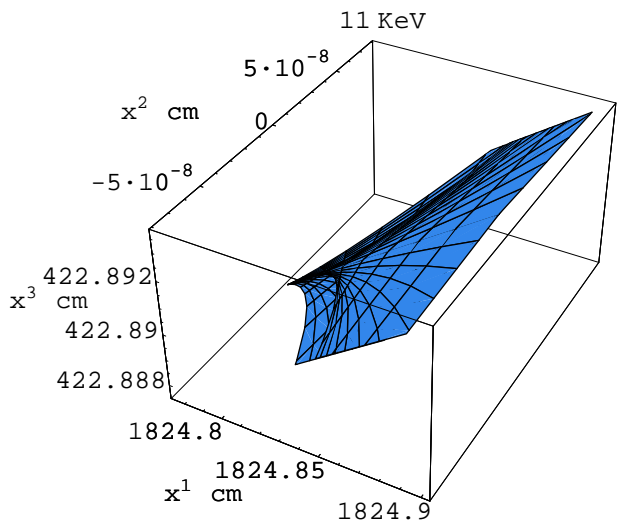}
\end{center}
\caption{ Caustic at $9.46\leq \protect\xi ^{1}\leq 9.475$.}
\label{fig7a.eps}
\end{figure}

\begin{figure}[tbp]
\begin{center}
\epsfig{figure=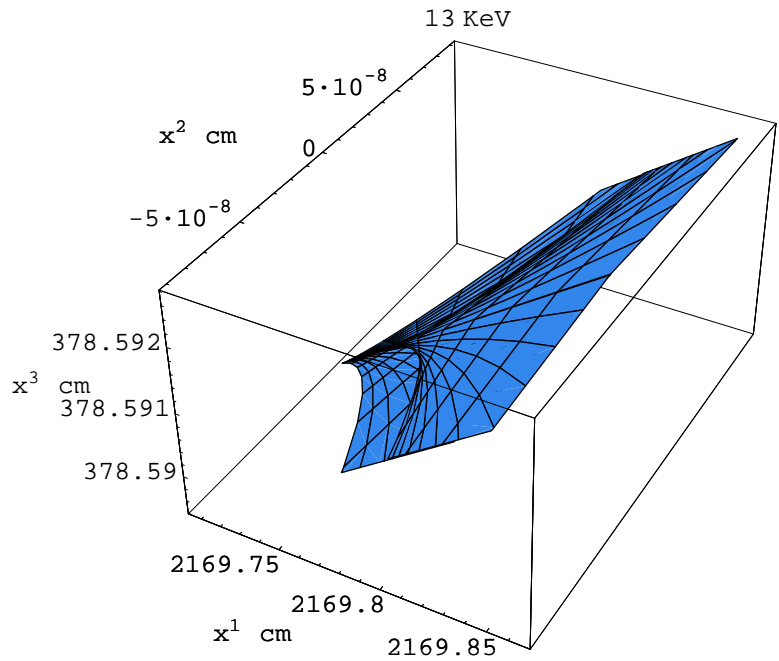}
\end{center}
\caption{Caustic at $3.994\leq \protect\xi ^{1}\leq 4.013$.}
\label{fig7b.eps}
\end{figure}

\begin{figure}[tbp]
\begin{center}
\epsfig{figure=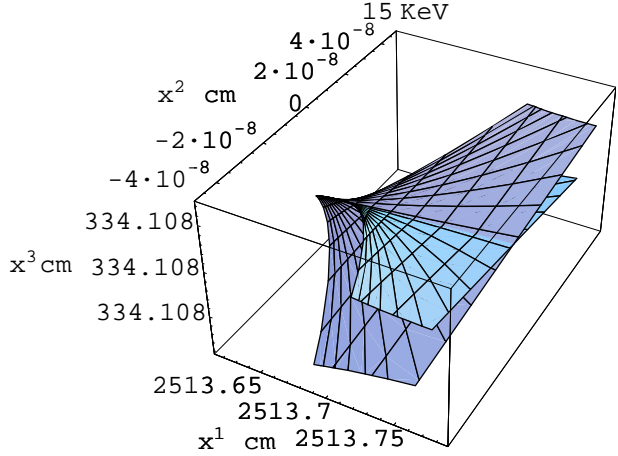}
\end{center}
\caption{ Caustic at $-0.007\leq \protect\xi ^{1}\leq 0.004$.}
\label{fig7c.eps}
\end{figure}

\begin{figure}[tbp]
\begin{center}
\epsfig{figure=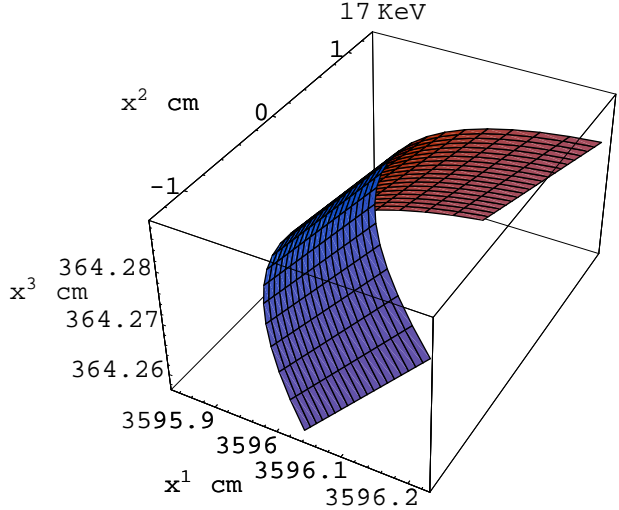}
\end{center}
\caption{ Caustic at $-3.069\leq \protect\xi ^{1}\leq -3.06$.}
\label{fig7d.eps}
\end{figure}

It is worth mentioning that even in the simplest case we deal with rather
complicated (that is clear from these diagrams) combination of elementary
differentiable peculiarities (catastrophes) $A_{1},A_{2},A_{3},A_{4}$ and $%
D_{4}^{\pm }$ (by V.I. Arnold classification) of phase function $\Phi _{(
{\bf l})}(\xi ^{1},\xi ^{2},\varepsilon )+S({\bf r},\xi ^{1},\xi ^{2})$.
That is why one can not even talk about the fit (frequently met in the X-ray
optics articles, see for ex. \cite{[21]}) of X-ray quanta intensity
distribution\ near the caustics (focus) with Gaussian function \cite{[20]} .
The caustics intensity distribution is described by special catastrophes
functions the simplest of them being Airy functions and Piersey integral %
\cite{[22]} investigated in details in articles \cite{[23]}. If one models
the intensity distribution by Gaussian beams, it is necessary to use the
special method of Gaussian beams summing discussed in \cite{[24]}.

\section{Acknowledgments}

We are indebted to Prof. S. Maksimenko and Prof. G. Slepyan for stimulating
discussions.

\end{document}